\documentclass[]{raa}            % referee version: for submission
\usepackage{graphicx,times}
\usepackage{natbib}

\providecommand{\bm}[1]{\mbox{\boldmath$#1$\unboldmath}}

\begin{document}

   \title{Multiwavelength study of nearly face-on low surface brightness disk galaxies
% $^*$
%\footnotetext{\small $*$ Supported by the National Natural Science Foundation of China.}
}

 \volnopage{ {\bf 2010} Vol.\ {\bf XX} No. {\bf XX}, 000--000}
   \setcounter{page}{1}

   \author{Dong Gao
      \inst{1,2,3}
   \and Yan-Chun Liang
      \inst{1,2}
   \and Shun-Fang Liu
      \inst{1,2,3}
   \and Guo-Hu Zhong
       \inst{1,2,3}
   \and Xiao-Yan Chen
        \inst{1,2,3}
   \and Yan-Bin Yang
        \inst{1,2,4}
   \and Francois Hammer
         \inst{4}
   \and	Guo-Chao Yang
        \inst{1,2,5}
   \and Li-Cai Deng
        \inst{1,2}
   \and Jing-Yao Hu
        \inst{1,2}
   }
%% Here is an example of three authors come from different institutes.
%% For single author or all the authors from an institute, use "\inst{}" only

   \institute{National Astronomical Observatories, Chinese Academy of Sciences, 
   A20 Datun Road, 100012 Beijing, China; {\it ycliang@nao.cas.cn, dgao@nao.cas.cn}\\
        \and 
            Key Laboratory of Optical Astronomy, National Astronomical 
            Observatories, Chinese Academy of Sciences, Beijing 100012, China 
        \and
             Graduate School of the Chinese Academy of Sciences, 100049 Beijing, China
	\and 
	     GEPI, Observatoire de Paris-Meudon, 92195 Meudon, France   \\   
	\and      
            Department of Physicals,Hebei Normal University, Shijiazhuang 050016, China \\
\vs \no
   {\small Received [2010] [x] [x]; accepted [2010] [x] [x] }
}

\abstract{We study the ages of a large sample (1,802) of nearly face-on disk 
low surface brightness 
galaxies (LSBGs) by using the evolutionary population synthesis (EPS) model PEGASE 
with exponential decreasing star formation rate
to fit their multiwavelength spectral energy distributions (SEDs)
 from far-ultraviolet (FUV) to 
 near-infrared (NIR). The derived ages of LSBGs are
 1-5 Gyr for most of the sample 
 no matter the constant or varying dust extinction is adopted, 
 which are similar to most of the previous studies on smaller samples. 
 This means that these LSBGs formed their majority of stars quite recently. 
 However, a small part of the sample ($\sim$2-3\%) have larger ages as 5-8 Gyr, 
 meaning their major star forming process may occur earlier.
 At the same time, a large sample (5,886) of  high surface brightness 
galaxies (HSBGs) are selected and studied in the same method for comparisons. 
The derived ages are 1-5 Gyr for most of the sample (97\%) as well.
These may mean that 
 probably these LSBGs have no much different star formation history 
 from their HSBGs counterparts. But we should notice that the HSBGs are about 0.2 Gyr younger generally,
 which could mean that the HSBGs may have more recent 
 star forming activities than the LSBGs.
}

   \authorrunning{D. Gao, Y. C. Liang, S. F. Liu et al. }            %author_head in even pages
   \titlerunning{Multiwavelength study of nearly face-on LSB disk galaxies }  % title_head in odd pages
   \maketitle

%% The author head (on even pages) and the title head (on odd pages) will be
%% automatically extracted from \author{} and \title{}. Whenever the title is too long,
%% you will be asked to supply a shorter one by inserting either \authorrunning{} or
%% \titlerunning{} before \maketitle. Anyway, you can specify your own heads.
%%
%%
%% Note: In the following text body of your manuscript, please note several differences from
%%       other major journals:
%% (1) \subsection{Please Capitalize the First Letter of Each Notional Word in Subsection Title}
%% (2) Please Capitalize the First Letter of Each Notional Word in all tables' captions

%
%________________________________________________ sections below
%
\section{Introduction}           %% first-level sections will be auto-capitalized
\label{sect:intro}

Low Surface Brightness Galaxies (LSBGs) 
were often hard to find owing to their faintness compared with the night sky, thus
their
properties were seldom studied and their contribution to the galaxy population were underestimated
for a long time. However,  it has been suggested that LSBGs may comprise up to 1/2 of the local galaxy
population (McGaugh, Bothun \& Schombert 1995). 

An initial quantitative study about LSBGs
was done by Freeman (1970), who noticed that the central surface brightness of their 28 out of 36 disc
galaxies fell within a rather narrow range, $\mu_{0}$(B)=21.65 $\pm$ 0.3 mag arcsec$^{-2}$. This could be
caused by selection effects  (Disney 1976; Zwicky 1957).
 Since then, many efforts have been made to search for more LSBGs from surveys (see the 
 reviews by Bothun, Impey, McGaugh 1997; Impey \& Bothun 1997;
 and the introduction part in Zhong et al. 2008). These include giant LSBGs (Sprayberry et al. 1995),
 red LSBGs (O'Neil et al. 1997) and the most common cases of ``blue LSBGs" as late-type, disk-dominated
 spirals with $\mu_{0}$(B)$>$ 22 mag arcsec$^{-2}$ (Zhong et al. 2008 and the references therein). 
 The LSBGs were generally thought to be unevolved systems with
 low metallicity (McGaugh 1994), low star formation rate (van der Hulst et al. 1993),
 a relatively high gas fraction (de Blok et al. 1996) 
 and large amounts of dark matter (de Blok \& McGaugh
 1997). The age of LSBG is also an important topic to study. 
  
 The ages of the LSBGs have long been a matter of controversy.  Broadband
photometric studies,  complemented by H$\alpha$ emission line data and
synthetic stellar population modeling, predict quite a wide range for the ages
of blue LSBGs: from 1-2 Gyr (Zackrisson et al. 2005) to 7-9 Gyr (Padoan et al.,
1997; Jimenez et al. 1998). Ronnback \& Bergvall (1994) carried out the
multicolor studies of a sample  of extremely blue galaxies with only  small
radial color gradients. In their work,  BVI photometry was interpreted using
spectral evolutionary models, and ages typically higher than 3 Gyr were found.
Using optical/near-infrared (NIR) broadband photometry together with H$\alpha$
emission line data for a subsample of this sample,  Zackrisson et al. (2005)
found that the current observations cannot rule out  the possibility that these
blue LSBGs formed as recently as 1-2 Gyr ago. 
 Schombert et al. (2001) argued
on the basis of their V-I colors and relative HI content that the most gas-rich
LSBGs should typically have mean stellar ages below 5 Gyr.  
Padoan et al. (1997) and Jimenez et al. (1998)  used
UBVRI photometry to conclude that most of the galaxies in their sample appeared
to be older than about 7 Gyr.  Vorobyov et al. (2009) used numerical
hydrodynamic modelling to study the long-term ($\sim$13 Gyr) dynamical and
chemical evolution of blue LSBGs adopting a sporadic scenario for star
forming. Their modelling strongly suggested the existence of a minimum age for
blue LSBGs: 1.5-3.0 Gyr (from model B-V colors and mean oxygen abundances), or
5-6 Gyr (from model H$\alpha$ equivalent widths). The later value may decrease
slightly if LSBGs host truncated initial mass functions (IMFs) with a smaller upper mass limit. 
Haberzettl, Bomans, Dettmar (2005) studied the star formation history of a
sample of 7 LSBGs in the HDF-S. The comparison of 
measured spectral energy distributions (SEDs) with the synthetic spectra
extracted from the synthesis evolution model PEGASE 
(Fioc \& Rocca-Volmerange 1997, assuming an exponentially
declining star formation rate) suggested the ages of the dominant stellar
population  between 2 to 5 Gyrs. This implies that the major star formation
event of LSBG galaxies took place at much later stages (at z$\sim$0.2 to 0.4).

  These studies show that the ages of LSBGs are in debate, and could
  spread in a wide range, although the working sample is small in these investigations.
  Surely much more efforts and much larger sample are needed for detailed studies 
  on the ages of LSBGs. 
  Moreover, most of these investigations are based on optical photometry which has
  obvious shortcoming in estimating ages of galaxies. As we know,
  the youngest stars may
 dominate the ultraviolet (UV) and optical part of the spectrum, and an old stellar component will thereby be extremely difficult to
 detect. With the inclusion of NIR
 broadband photometry one would expect to be able to constrain the ages more strictly 
 (Bell et al. 2000; Zackrisson et al. 2005). The UV data are also needed to 
 build SEDs of galaxies in a wider range, which will help to 
 derive ages of galaxies more reliably.
  At present, the modern digital sky survey in large sky coverage
 provide us much larger sample of galaxies for doing detailed and statistic studies 
 on many fields. Moreover, the multiwavelength
 data of a large set of astronomical objects from far-ultraviolet (FUV) to 
 NIR have been carried 
 by high efficient surveys and have been released for public use, such as 
 Galaxy Evolution Explorer (GALEX) in FUV and near-ultraviolet (NUV), Sloan Digital Sky Survey (SDSS) in optical
 and Two Micron All Sky Survey (2MASS) in NIR. These multiwavelength data will help us to well study
 the properties of a large sample of LSBGs. We focus on the ages of LSBGs in this work.   

We will study the ages of a large sample of nearly face-on disc LSBGs selected from SDSS
Data Release 7 (DR7, Abazajian et al. 2009) 
main galaxy sample (MGS, Strauss et al. 2002),
and then matched them with the 2MASS /NIR and GALEX/UV data, so that their SEDs could cover from 
FUV to Ks (1350\AA~ to 2.17$\mu$m) bands. The selected sample of LSBGs are 1,802
with their $\mu_{0}$(B) $\geq$ 22 mag arcsec$^{-2}$. 
Indeed, this is a follow-up work of 
Zhong et al. (2008) and Liang et al. (2010).
In Zhong et al. (2008), we selected a la
SDSS Data Release 4 (DR4, Adelman-McCarthy et al. 2004)
main galaxy sample (Strauss et al. 2002), and presented their basic photometric properties
including correlations of disk scalelength versus  
B-band absolute magnitude and
distance, stellar populations from colors. 
In Liang et al. (2010), we studied the spectroscopic properties of this large sample 
of LSBGs, including dust extinction, strong emission-line ratios,
metallicities and stellar mass-metallicity relations.
In this work, we firstly extend the sample from DR4 to DR7, 
then try to obtain their multiwavelength data from FUV to Ks 
based on the public survey data
from GALEX GR4/GR5 and 2MASS Extended Source Catalogue (XSC). Then, we use the 
evolutionary population synthesis (EPS) model
PEGASE to fit their multiwavelength SEDs from FUV to NIR, and derive the ages of these galaxies.
At the same  time,
a large sample of nearly face-on disk high surface brightness 
galaxies (HSBGs, 5,886) 
 with $\mu_{0}$(B) $<$ 22 mag arcsec$^{-2}$ are selected and applied to same analyses for comparisons.

   This paper is organized as follows. In Sect.\,2, 
    we describe the multiwavelength observations and the selected sample. 
    In Sect.\,3 we describe the EPS model PEGASE, which is what we used to fit the SEDs of
    the galaxies,
    and the adopted parameters. In Sect.\,4, we present the fitting method
    and the derived ages of the sample galaxies.
    In Sect.\,5, the incompleteness effects are discussed.
    The discussions are given in Sect.\,6. 
    We summarize and conclude the work in Sect.\,7.
Throughout this paper, a cosmological model
with  $H_0$=70 km s$^{-1}$ Mpc$^{-1}$, $\Omega _M$=0.3 and $\Omega _\Lambda =0.7$
has been adopted. All the magnitudes are in AB system.

\section{The sample}
\label{sect:Obs}

Our sample galaxies are selected by matching three databases from multiwavelength 
surveys, i.e., the SDSS for 
ugriz optical, 2MASS for JHKs NIR, and GALEX for FUV and NUV bands.  

\subsection{The SDSS sample}

The SDSS\footnote{http://www.sdss.org} is the most
ambitious astronomical survey ever undertaken in imaging and
spectroscopy (York et al. 2000;
Stoughton et al. 2002). The photometric and spectroscopic
observations were conducted using the 2.5-m SDSS telescope at the Apache Point Observatory in New Mexico,
USA. The imaging data were done in
drift-scan mode in ugriz five bands, with effective central wavelengths of 3551, 4686,
6166, 7480 and 8932 \AA, respectively (Gunn et al. 1998), and
the 95\% completeness limits for point sources are 22.0, 22.2, 22.2, 21.3
and 20.5 mag, respectively.  The spectra are flux- and
wavelength-calibrated with 4096 pixels from 3800 to 9200 \AA~ at
$R\sim$1800 (Stoughton et al. 2002). 

 We select 21,666 nearly face-on disk LSBGs from the SDSS-DR7 MGS 
 by following the criteria used in 
 Zhong et al. (2008).  These are  $fracDev_r$ $<$ 0.25 (indicating the fraction of luminosity
contributed by the de Vaucouleurs profile relative to exponential
profile in the $r$-band is much small), $b/a$ $>$ 0.75 (for nearly face-on ones,
$a$ and $b$ are the semi-major and
semi-minor axes of the fitted exponential disk, respectively), $M_B$ $<$ -18 (excluding 
few dwarf galaxies with this B-band absolute magnitude cut)
 and $\mu_0$(B)$\geq$ 22 mag arcsec$^{-2}$.  
 These selection criteria have also been described in Zhong et al. (2010).
 At the same time, 
 30,896 nearly face-on disk HSBGs with $\mu_0$(B)$<$ 22 mag arcsec$^{-2}$ are selected for comparisons. 
  The central surface brightnesses of our sample galaxies are 
 calculated by using the corresponding parameters provided in the SDSS
 catalog following the method given in Sect.2.2 and Eq.(6) of Zhong et al. (2008). 
 We use Petrosian magnitudes (and their errors) to represent the optical brightness of the SDSS galaxies.
 The Petrosian magnitudes could recover essentially all of the flux of an exponential galaxy profile
 (Stoughton et al. 2002), which are just the case of our sample here.  
 All the magnitudes are converted to be in AB system following Hewett et al. (2006).

\subsection{The SDSS-2MASS matched sample}

   We use the 2MASS Extended Source Catalogue (XSC)\footnote{http://www.ipac.caltech.edu/2mass/} 
   to match with the SDSS selected sample. Using the Mt. Hopkins northern 1.3-m
telescope and the CTIO southern 1.3-m telescope in Chile, the 2MASS covered almost entire sky in the
J(1.25$\mu$m), H(1.65$\mu$m) and Ks(2.17$\mu$m) bands, with spatial resolution of 3$^{\prime \prime }$. 
The Point-source sensitivity limits (10$\sigma$) are 15.8 (0.8 mJy), 15.1 (1.0 mJy),
and 14.3 (1.4 mJy) mag at J, H, and Ks, respectively. The extended source sensitivity (10$\sigma$)
is $\sim$1 mag brighter than the point-source limits, or 14.7 (2.1 mJy), 13.9 (3.0 mJy),
and 13.1 (4.1 mJy) mag at J, H, and Ks, respectively, 
with the precise threshold depending on the
brightness profile of the objects (Jarrett et al. 2000). 
The magnitude limit of typical extended sources in the
Ks band is about 15.3 AB mag (Lee et al. 2010). 

By matching our SDSS sample galaxies with the 2MASS XSC catalog within 3$^{\prime \prime }$
following Blanton et al. (2005) and Lee et al. (2010), 
we obtain 3,523 LSBGs and 9,483 HSBGs.
For the JHKs magnitudes (and their errors) of the galaxies,
we adopt those from fit extrapolation (j,h,k-m-ext) provided by the 2MASS 
database\footnote{see http://irsa.ipac.caltech.edu/applications/Gator/}, 
which are assumed
to better trace the galactic luminosity (Jarrett et al. 2000; Blanton et al. 2005).
All the magnitudes are converted to be in AB system following Blanton et al. (2005).

\subsection{The SDSS-2MASS-GALEX matched sample}

   We use the GALEX (Martin et al. 2005) survey GR4/GR5\footnote{http://galex.stsci.edu/GR4/} 
   data to match with our selected sample from SDSS and 2MASS. 
   The GALEX was
launched in 2003 April, and will cover the whole sky in the FUV (1350 - 1750 \AA) and 
NUV (1750 - 2800 \AA) bands, with spatial resolutions of 6.0$^{\prime \prime }$
in the FUV and 4.5$^{\prime \prime }$ in the NUV. The central wavelengths of FUV and NUV
are  1528\AA~ and 2271\AA, respectively.
In this paper, only the GALEX
objects detected both in the NUV and FUV bands were used.
The GALEX data products are made available to the general public 
via the MultiMission Archive at Space Telescope Science Institute (MAST).
The GALEX survey design was done for five modes, in which two of them are more related to
our work here, i.e., the
All-sky Imaging Survey (AIS) with the goal to survey the entire sky subject to a sensitivity of 
$m_{AB}$$\sim$ 20.5, and the Medium Imaging Survey (MIS), which covers 1000 deg$^2$  with extensive overlap of the
SDSS (Martin et al. 2005; Morrissey et al. 2007).

We use the CASJobs\footnote{http://mastweb.stsci.edu/gcasjobs/} 
to match our SDSS-2MASS sample with the GALEX GR4/GR5 database by importing coordinate list
and then obtain their magnitudes in FUV and NUV 
(see the help in website\footnote{http://galex.stsci.edu/doc/CASJobsXTutorial.htm}). 
Firstly, we import a prepared coordinate list into the database,
 which is for the RA and DEC of our SDSS-2MASS matched sample.   
Then a new table of this is created in the database.
Next we search for neighbors for the objects within a radius of 5$^{\prime \prime }$, 
and then obtain a new table of GALEX-matched objects which provide 
the unique GALEX object identifier (objid) 
as matched\_id, and also the search\_id to mark which original 
object it corresponds to. 
In the case of multiple matches, we choose the nearest one with minimum matching distance.
Finally, we obtain the magnitudes, fluxes and their errors for these objects.
We use the GALEX products given in the ``-mcat.fits" file, which is the Merged (FUV+NUV) source catalog.
We adopt ``nuv\_mag" and ``fuv\_mag" (and their errors) as the magnitudes of our matched objects,
which are the NUV calibrated magnitude and FUV calibrated magnitude,
respectively,
and ``calibrated" means that values have been converted to AB magnitudes (see help in 
the pipeline data 
guide\footnote{http://galexgi.gsfc.nasa.gov/docs/galex/Documents/GALEXPipelineDataGuide.pdf}).

\begin{figure} 
\centering
\includegraphics [width=5.5cm, height=5.5cm]{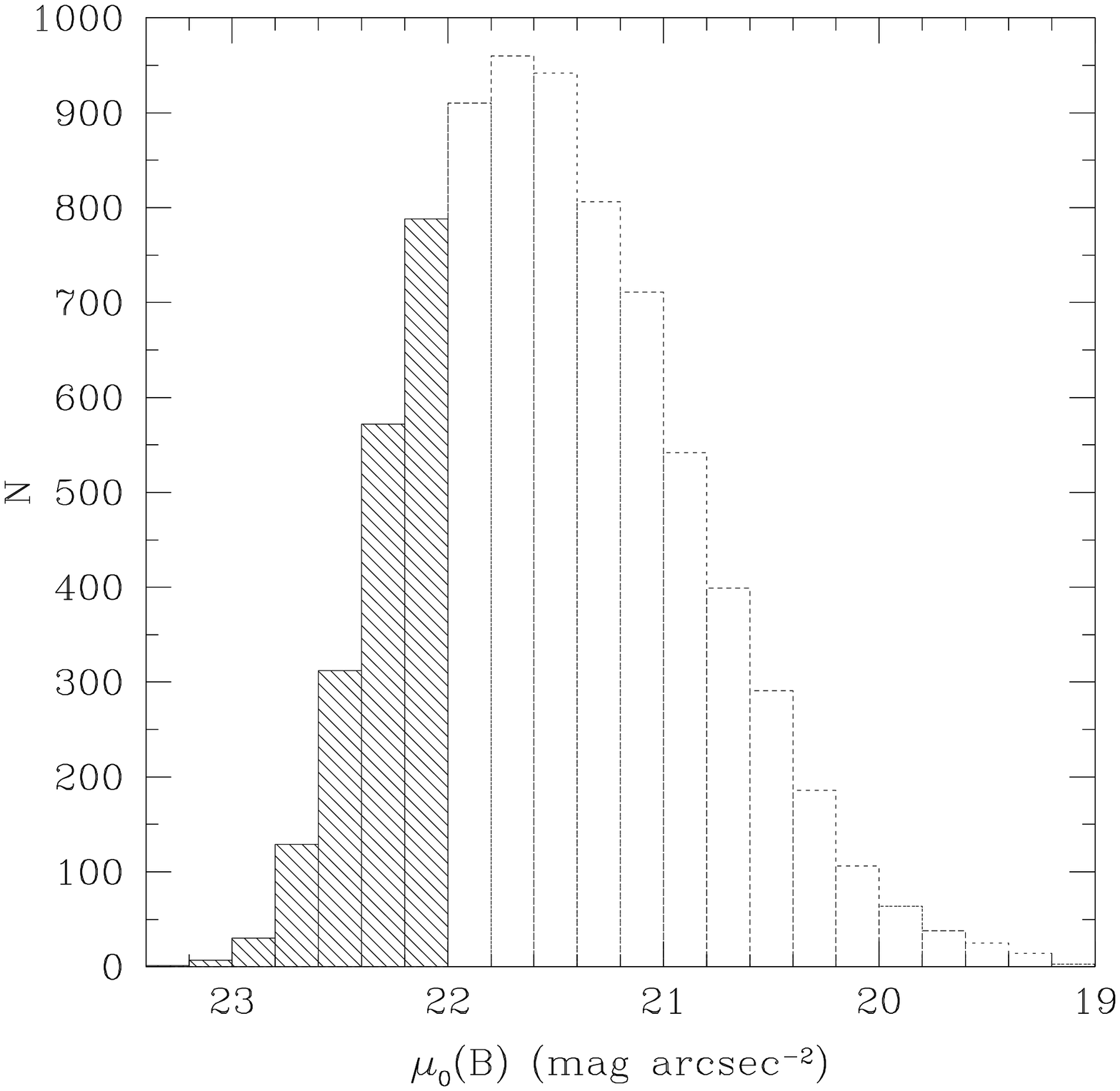} 
\includegraphics [width=5.5cm, height=5.5cm]{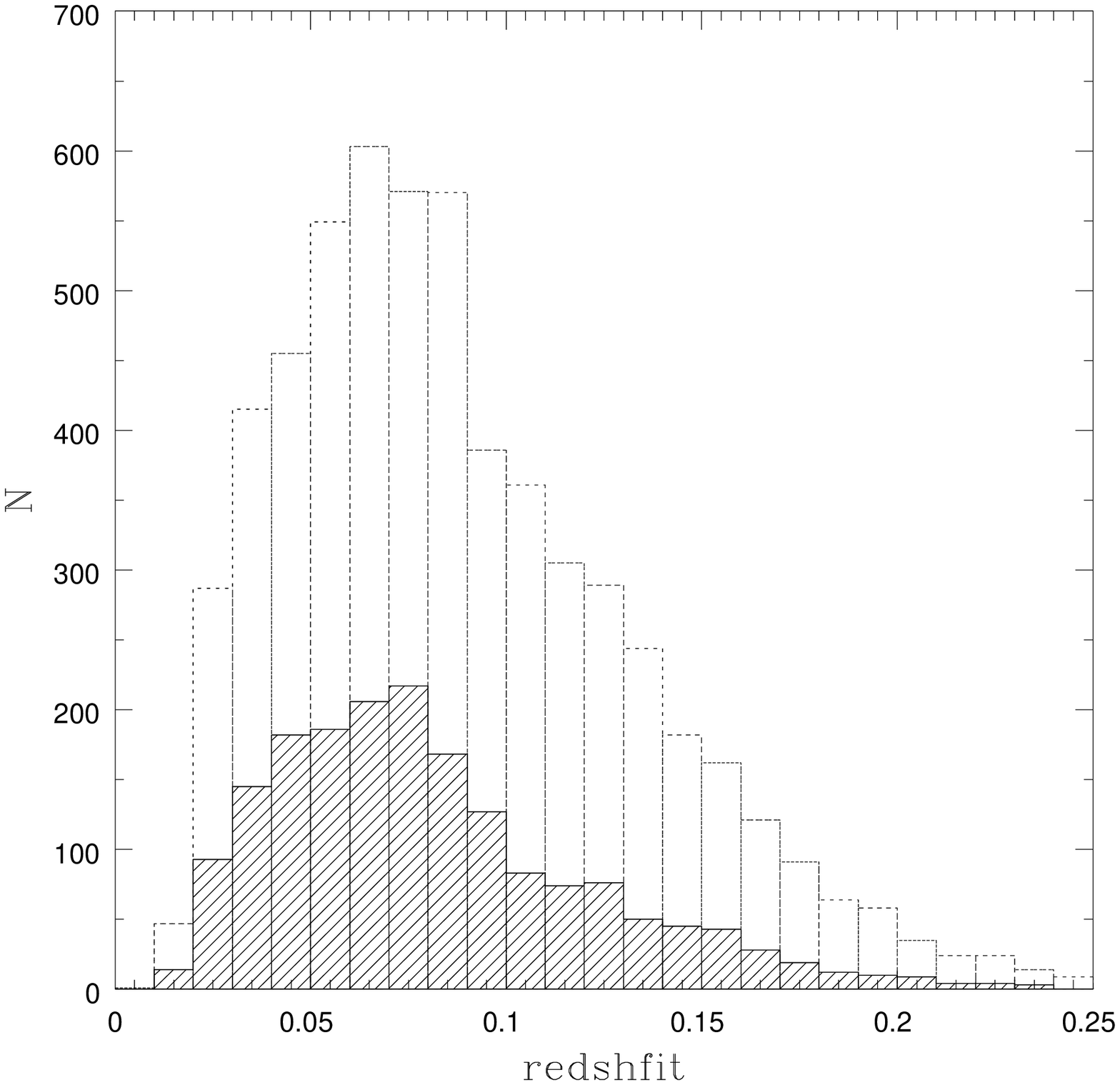} 
\caption {Histogram distributions of $\mu_{0}$(B) (with bin of 0.1) and redshift 
(with bin of 0.01) 
for our sample galaxies with FUV-to-NIR
multiwavelength observations. The shaded regions refer to the 
LSBGs with $\mu_{0}$(B) $\geq$ 22 mag arcsec$^{-2}$, 
and the dashed-lines refer to the HSBGs. }
\label{his.mu.z}
\end{figure}

Finally, we select 1,802 LSBGs and 5,886 HSBGs with FUV-to-NIR
 multiwavelength observations.
The histogram distributions of their $\mu_{0}$(B) and redshift are given in Fig.~\ref{his.mu.z}.
The median values of $\mu_{0}$(B) are 22.24 mag arcsec$^{-2}$ for LSBGs and 
21.35 mag arcsec$^{-2}$ for HSBGs 
(mean as 22.28 and 21.25), respectively.
The corresponding median values of redshifts are 0.0732 and 0.0801 
(mean as 0.0810 and 0.0886), respectively.

\subsection{Corrections on magnitudes}

The effects of foreground Galactic extinction on the observed magnitudes 
were calculated by using the reddening
maps of Schlegel, Finkbeiner \& Davis (1998). The corresponding values 
of the SDSS and 2MASS magnitudes have been provided by Blanton et al. (2003)
at the NYU-VAGC\footnote{http://sdss.physics.nyu.edu/vagc/} catalog (thanks to them). For the GALEX FUV and NUV magnitudes, 
the corrections for interstellar extinctions were calculated using
$A(FUV)=8.16E(B-V)$ and $A(NUV)=8.90E(B-V)$ (Rey et al. 2007; Kinman et al. 2007),
and the $E(B-V)$ were taken from Schlegel, Finkbeiner \& Davis (1998).

   The magnitudes in our sample are then corrected for the redshifts.
   We calculated the K-corrections  
   using their K\_CORRECT program, version 4.1.4, originally developed 
   by Blanton et al. (2003) and now extended to handle GALEX data (Blanton \& Roweis 2007).
 We corrected the
observed magnitudes (FUV-to-NIR magnitudes) of the sample galaxies into the magnitudes at redshift z = 0. 
We did not consider the evolutionary correction on the magnitudes of galaxies since it won't be
large in the redshift range of our sample. We did not further consider aperture corrections
for the magnitudes among three surveys since the Petrosian magnitudes in SDSS, 
fit extrapolation  magnitudes in 2MASS and calibrated magnitudes in GALEX have 
been good tracers of the total luminosities of these galaxies. 

For the internal reddening of the galaxy, we input dust extinction  
in the PEGASE model, then get the model SEDs
after extinction when do SED fittings. As for the values of 
internal extinction $A_V$, we
adopt two methods, one is constant dust extinction, i.e., $A_V$=0.6
mag for LSBGs and $A_V$=0.8 mag for HSBGs, and another one is 
assuming $A_V$ as a free
parameter varying from 0.1 to 1.6 in the fitting procedures,
which come from Liang et al. (2010) about the dust extinction estimates
from Balmer decrement for related samples.  More information
will be given in Sect.\,4.

\section{Evolutionary Synthesis Model PEGASE}
\label{sect:data}

Evolutionary population synthesis is a powerful tool of interpretation of the integrated
spectrophotometric observations of galaxies. The most common method of model-observation comparison for
stellar population analysis in galaxies or star clusters is SED fitting, 
with either least-squares or chi-squared minimization technique (Kong et al. 2000; 
Gavazzi et al. 2002; Jiang et al. 2003; Li et al. 2004a,b; 
Fan et al. 2006; Ma et al. 2002a,b, 2007, 2009a,b; Wang et al. 2010).   
There are several EPS models popularly used in astrophysical studies 
(Chen et al. 2009, 2010 for references therein). We use PEGASE in this work.

   PEGASE is an evolutionary spectral synthesis model for starbursts and
evolved galaxies of the Hubble sequence. It is continuous over an exceptionally
large wavelength range from 220 \AA~ up to 5$\mu$m. It was extended to the NIR
of the atlas of synthetic spectra of Rocca-Volmerange \& Guiderdoni (1988) with
a revised stellar library including cold star parameters and stellar tracks
extended to the thermally-pulsing regime of the 
asymptotic giant branch (TP-AGB) and the post-AGB phase. The synthetic stellar spectral library is
taken from Kurucz (1992), modified by Lejeune et al. (1997) to fit the observed
colors. A set of reference synthetic spectra at z = 0, to which the
cosmological k- and evolution e- corrections for high-redshift galaxies are
applied, is built from fits of observational templates (Fioc \&
Rocca-Volmerange 1997).

With the PEGASE code, we can compute the stellar SEDs of starbursts and evolved galaxies of the Hubble
sequence at any stage of evolution, within the metallicity range Z = 10$^{-4}$ to 10$^{-1}$. 
Typical parameters of PEGASE are the star formation rate (SFR) and IMF.
Assuming a standard IMF,
SFR and other initial conditions, such as dust extinction, 
the PEGASE code will give an
evolutionary history and some other important properties of a galaxy (Li et al. 2004a,b).

In our fitting analysis, 
the internal dust extinctions are considered in two methods, 
the constant A$_V$ (0.6 for LSBGs and 0.8 for HSBGs) and varying $A_V$ within A$_V$ = 0.1 - 1.6.
A zero initial metallicity of interstellar medium
(ISM) is taken.
 We adopt the exponential decreasing SFR ({\it SFR(t)} $\propto$~{\it e}$^{-t/\tau}$)
with $\tau$ = 0.1 - 15 Gyr varying in the fittings.
The IMF is assumed to follow the Salpeter (1955)
form, $\Phi$(M) = A$\times$ $M^{-\alpha}$  with ${\alpha}$ = 2.35 and a lower cutoff of $M_{l}$ = 0.1
$M_{\odot}$ and an upper cutoff $M_{u}$ = 125 M$_{\odot}$(Sawicki \& Yee 1998). 
 As a result, a  rest-frame modeled spectra
with various star formation histories are generated by running the
PEGASE code.

\section{Model Fits And Results}

We present the fitting method in Sect.~\ref{sect:analysis0},
and the fitting results in Sects.~\ref{sect:analysis1}, \ref{sect:analysis2}.
Extinction
affects intrinsic colors of the objects and hence accurate ages, so the photometric measurements must be
dereddened before running the fitting procedure. 
We use two methods for considering the internal dust extinction of the galaxies
in fitting procedures, the constant and varying $A_V$.

\subsection {The fitting method}
\label{sect:analysis0}

To estimate the ages of galaxies accurately, we add UV and NIR photometric data
points to the optical data.
As Ma et al. (2009b) discussed, Kaviraj et al. (2007) showed that the combination
of FUV and NUV photometry with optical observations in the standard broad bands
enables one to efficiently break the age-metallicity degeneracy.
The optical broadband colors have the more obvious problem
of age-metallicity degeneracy (Worthey 1994; MacArthur et al. 2004).
Again, de Jong (1996) showed that such degeneracy can be partially broken by
adding NIR photometry to optical colors, which has also been stated by
Bell et al. (2000), Wu et al. (2005) and Li et al. (2007). 

Our observational data consist of integrated luminosities 
through a given set of filters (FUV, NUV, ugriz, JHKs), 
thus we convolved the theoretical 
SEDs with these filter response curves to obtain 
synthetic ultraviolet, optical and NIR photometry for comparison.
The synthetic magnitude in the AB magnitude system for the $i$th filter
 can be computed as 
 \begin{equation}
 m_i=-2.5{\rm log}\frac{\int_{\lambda} F_\lambda\varphi_i(\lambda)d\lambda}{\int_{\lambda}\varphi_i(\lambda)d\lambda}-48.60,
 \end{equation}
 where $F_{\lambda}$ is the theoretical SED and $\varphi_i$ is the response curve
 of the $i$th filter of the sets of filter we used.
 
 We use a $\chi^2$ minimization test to examine which model SEDs are
 most comparable with the observed SEDs following
 \begin{equation}
\chi^2=\frac{1}{d}\sum^{10}_{i=1}\frac{[m^{obs}_{\lambda_i}-m^{mod}_{\lambda_i}(t)]^2}{\sigma^2_{i}},
 \end{equation}
where $m^{mod}_{\lambda_i}(t)$ is the integrated magnitude in the $i$th filter 
of a theoretical SED at age $t$,
$m^{obs}_{\lambda_i}(t)$ represents the observed integrated magnitude in the 
same filter, and $\sigma_i^2$ is the observational uncertainty for the
$i$th filter magnitude, $d$ is the number of degrees of freedom. 
We did not consider the uncertainty associated with the model itself here, which 
should be insignificant and not affect much our results.
More details about the fitting method can be referred to Li et al. (2004a,b) 
and Ma et al. (2009a,b).
The derived ages of our sample galaxies from such SED fitting analyses are given 
in next two subsections.

\subsection{Constant dust extinction}
 \label{sect:analysis1}

 We adopt constant $A_V$=0.6 mag for LSBGs and $A_V$=0.8 mag for HSBGs in fittings  
 by following Liang et al. (2010).
 In Liang et al. (2010), we derive
the $A_V$ value for each object from their spectroscopic Balmer decrement H$\alpha$/H$\beta$
(see their Fig.2 for histogram distributions).
The median values of $A_V$ of the four subsamples 
(with $\mu_0(B)$ in units of mag arcsec$^{-2}$:
 vLSBGs with 22.75-24.5, 
 iLSBGs with 22.0-22.75,
 iHSBGs with 21.25-22.0,
 vHSBGs with $<$21.25)
are 0.46, 0.63, 0.76, 0.83, respectively.
In our present work, most LSBGs have $\mu_0(B)$ within 22-23 mag arcsec$^{-2}$,
thus $A_V$=0.6 is acceptable to be their dust extinction. 
Also $A_V$=0.8 is reasonable to be the dust extinction of HSBGs.

\begin{figure}
\centering 
\includegraphics [width=4.8cm, height=4.8cm]{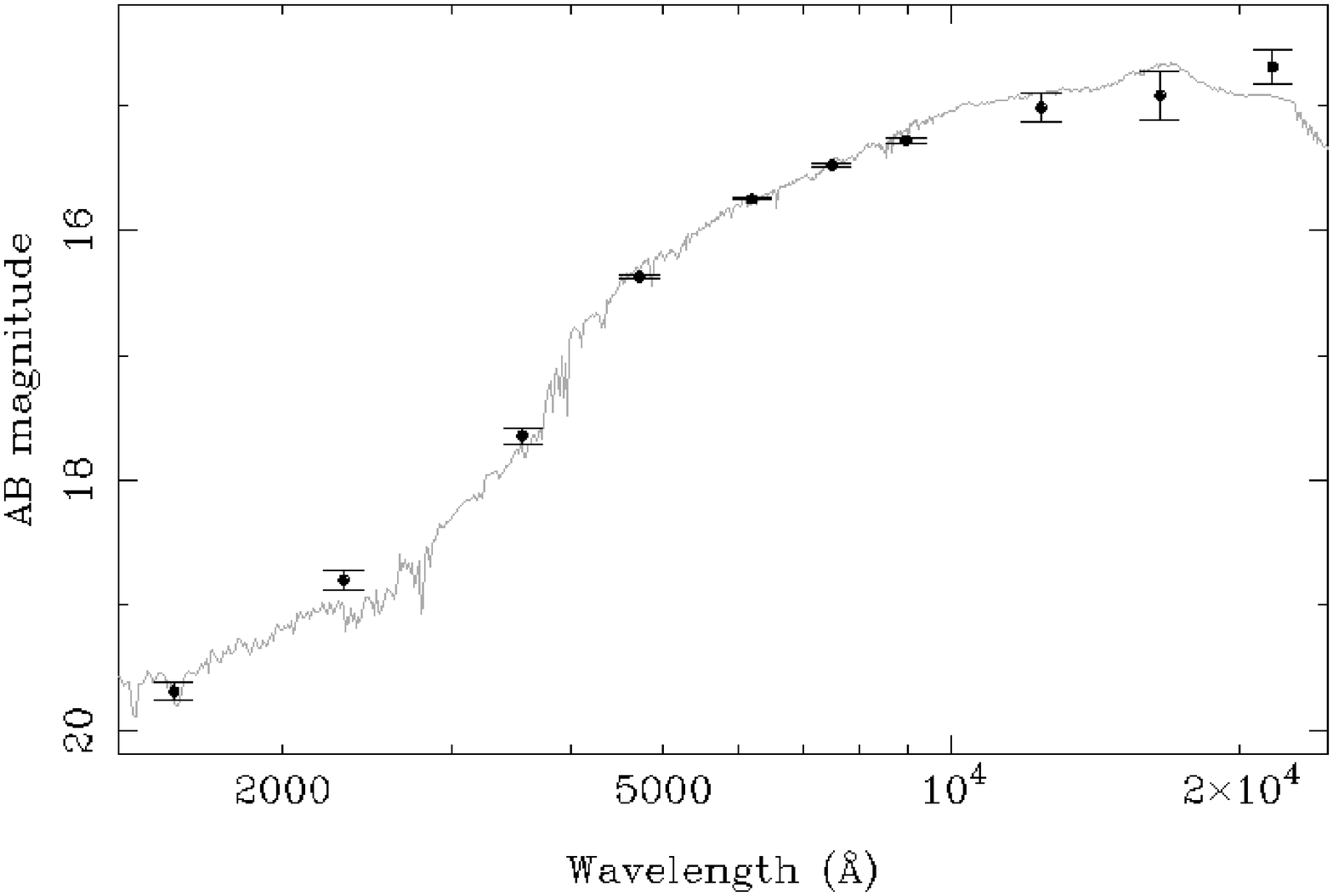} \hspace{-0cm}
\includegraphics [width=4.8cm, height=4.8cm]{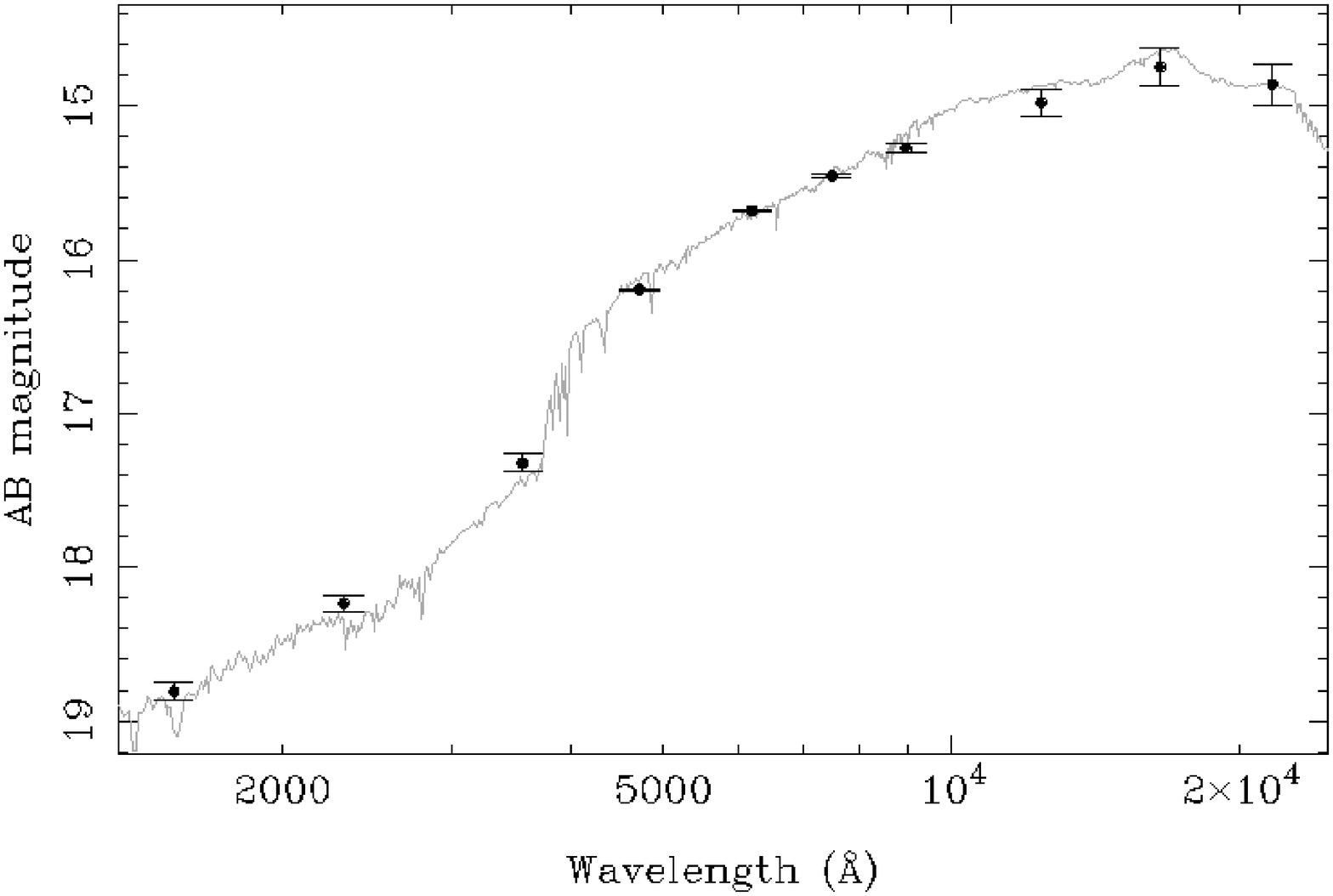}  \hspace{-0cm}
\includegraphics [width=4.6cm, height=4.2cm]{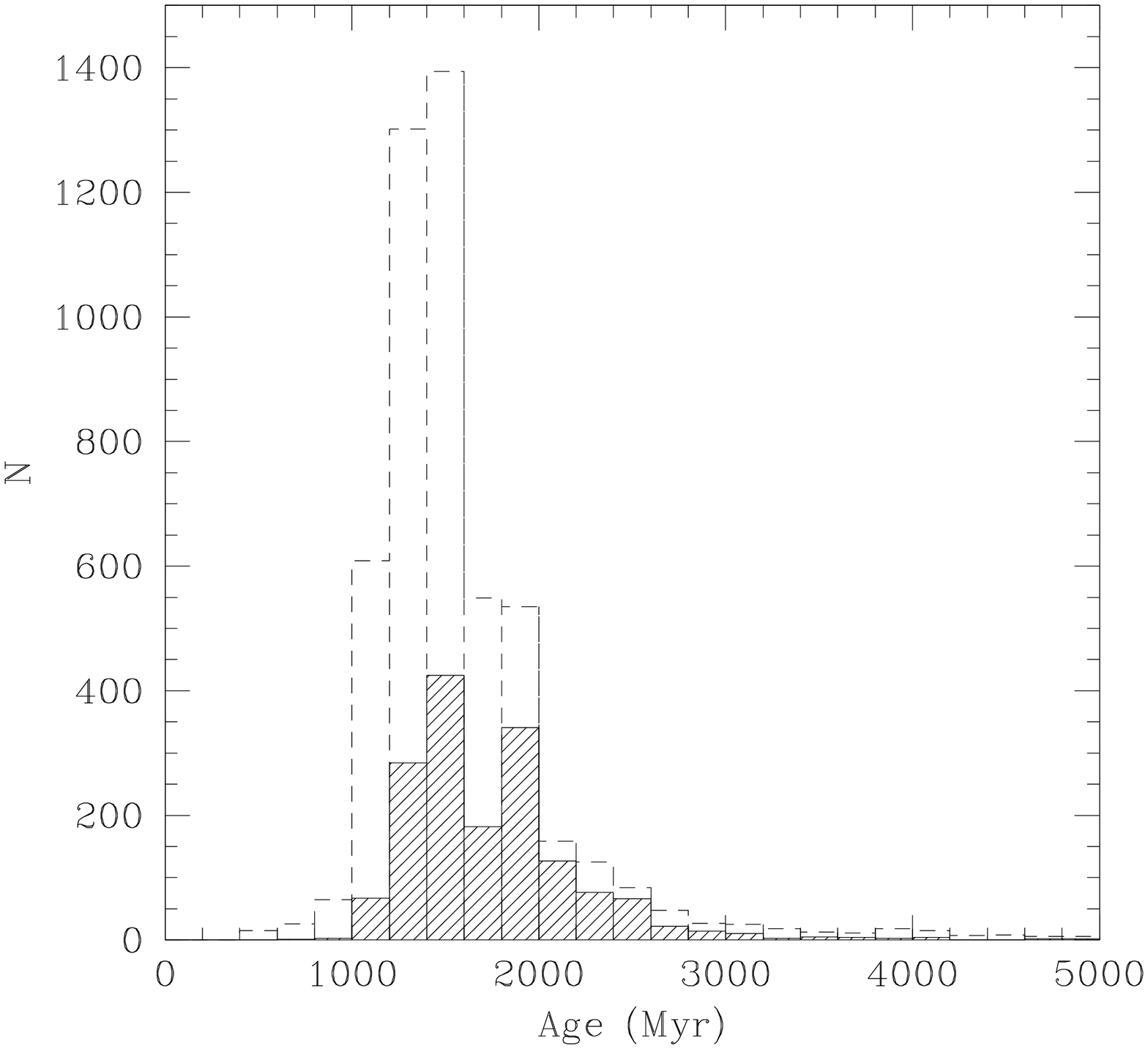} 
\caption{The fitting results with PEGASE model and constant dust extinction ($A_V$=0.6 for LSBGs
and $A_V$=0.8 for HSBGs) for our sample galaxies. The left panel shows the SED fitting 
for one example of LSBGs 
(RA=8.06543, DEC=0.90616 in 2000 epoch, the derived age is 1.57 Gyr), 
the middle panel shows the SED fitting for
one example of HSBGs (RA=6.75941, DEC=-9.63505  in 2000 epoch, the derived age is 1.44 Gyr)), 
and the right panel shows the histogram
distributions of the derived ages of LSBGs (shade region) and HSBGs (dashed-line)
with bin of 0.2 Gyr.
}
\label{Fig2}
\end{figure}

Figure~\ref{Fig2} shows the resulted SED fittings (left panel for one example of LSBGs;
middle panel for one example of HSBGs) and 
the histogram distributions of the derived ages of the galaxies (the right panel, where
the shaded region is for LSBGs and the dashed-line is for HSBGs).
The right panel shows that the ages of most of the LSBGs are 1-5 Gyr with median (mean)
value of 1.63 (1.75) Gyr. This means that the majority of their stars 
were formed quite recently. However, about 3\% of the LSBGs have larger ages,
5-8 Gyr. This small part of galaxies could probably form their stars at earlier time.
Similarly, the ages of most of HSBGs are 1-5 Gyr as well with the median value of 
1.47 (mean as 1.59) Gyr.
About 3\% of them have larger ages as 5-8 Gyr.
This similarity may mean that the LSBGs and HSBGs 
could have no much different star formation history.
This result is consistent with the metallicity estimates of similar samples
presented in Liang et al. (2010).
But we should notice the fact that the HSBGs
are slightly younger than the LSBGs, $\sim$0.2 Gyr shown by the median (mean) ages of the samples. 
{The median (mean) ages of the sample galaxies are 
presented in Table~\ref{tab1}.}

\subsection{Varying dust extinction} 
\label{sect:analysis2}

Now we adopt $A_V$ as a varying parameter 
from 0.1 to 1.6 in the fittings. 
This range of $A_V$ values is reasonable for LSBGs and HSBGs by following
Liang et al. (2010), and 1.6 is almost the upper limit of the sample
(also see Liang et al. 2007 for the SDSS star forming galaxies). 

\begin{figure}
\includegraphics [width=4.8cm, height=4.8cm]{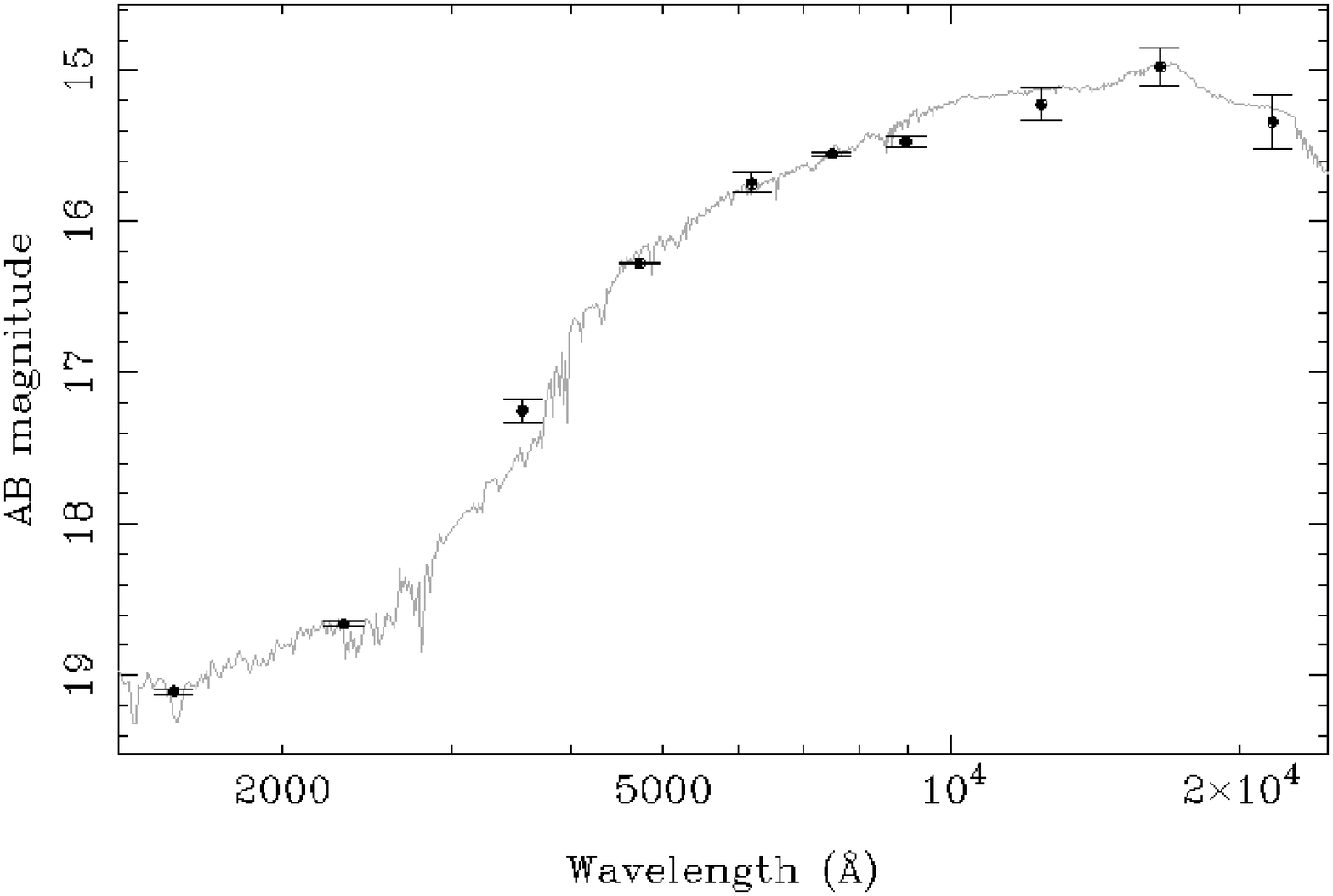} 
\includegraphics [width=4.8cm, height=4.8cm]{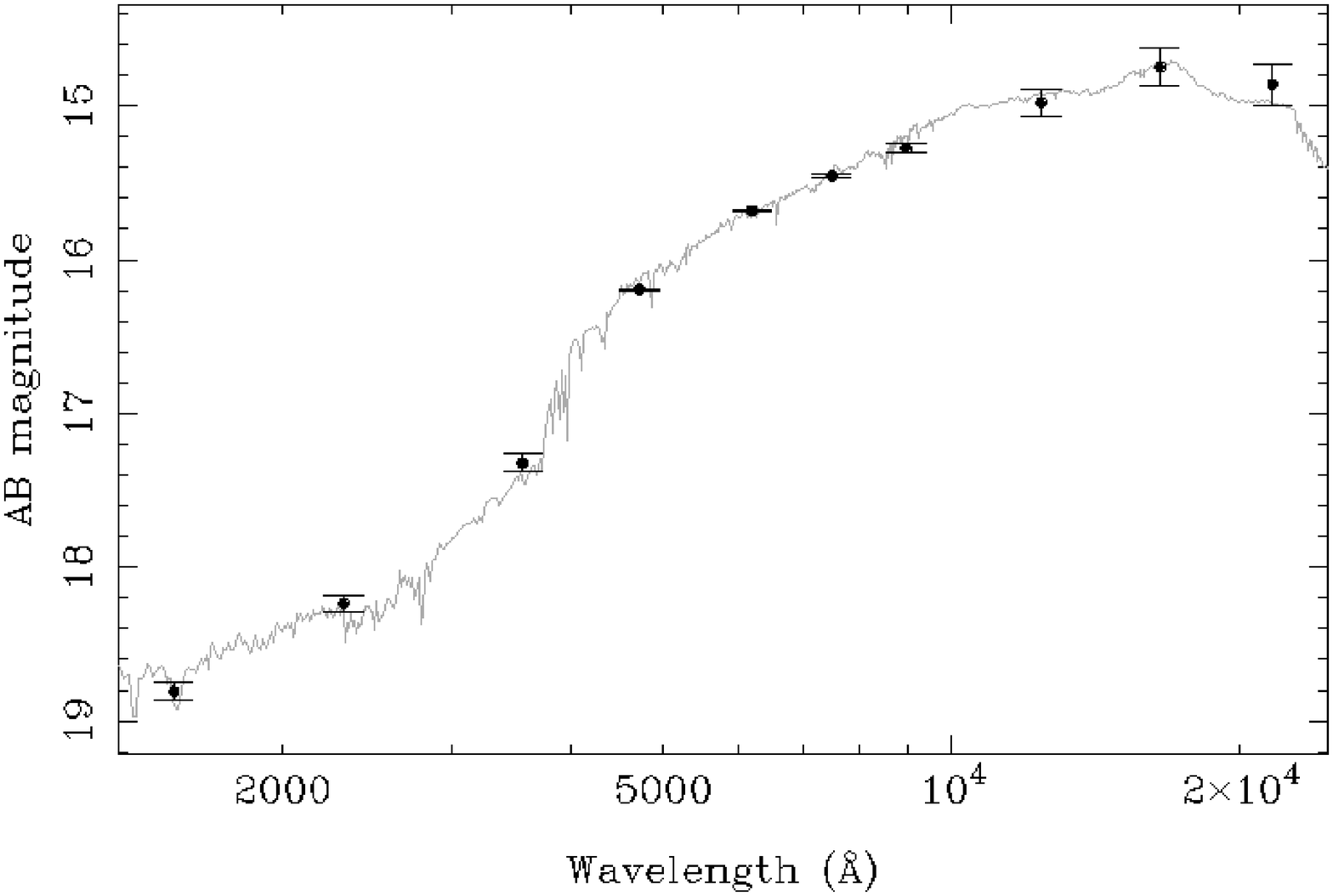} 
\includegraphics [width=4.6cm, height=4.2cm]{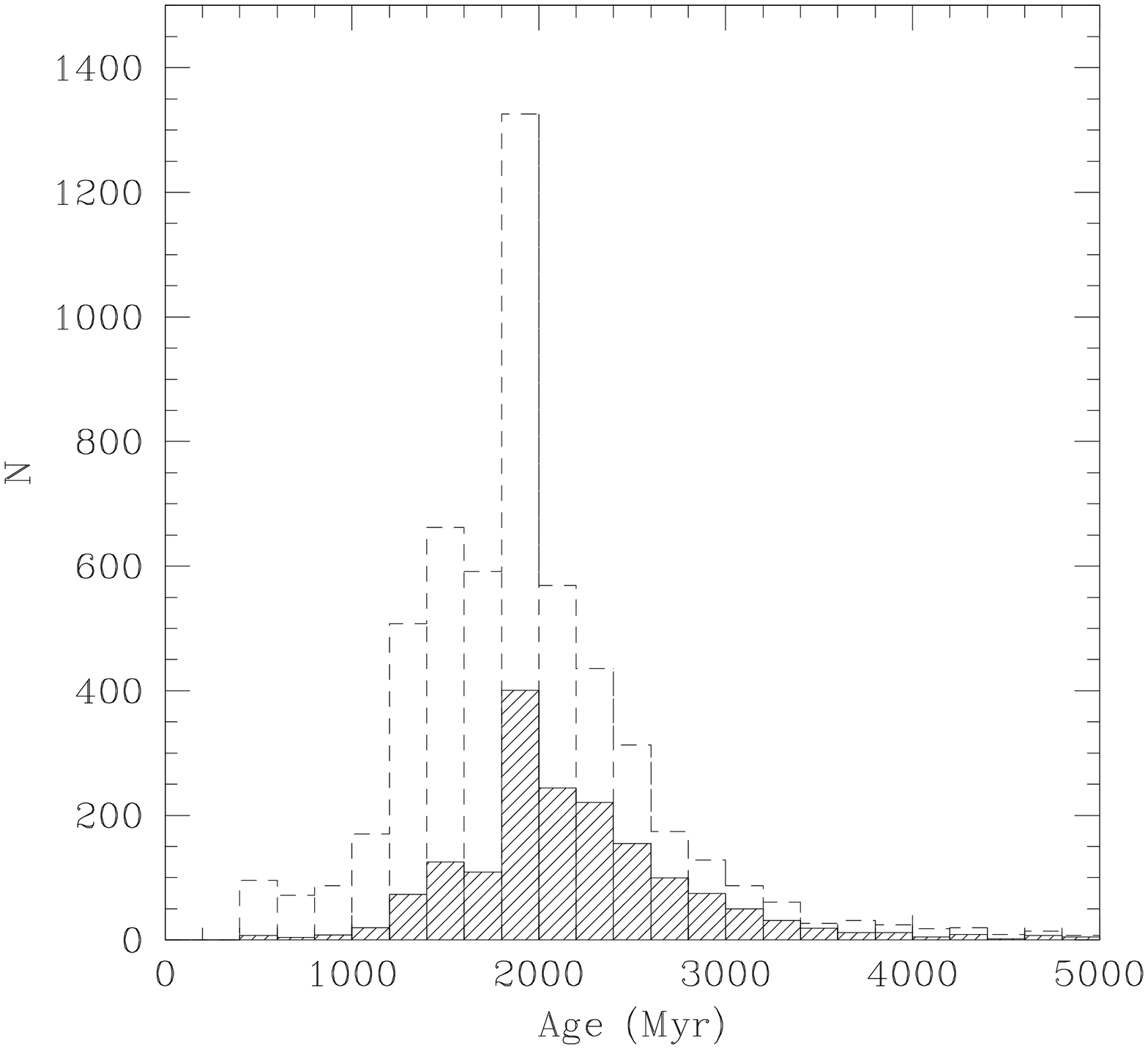} 
\caption{The fitting results with PEGASE model and varying dust extinction ($A_V$=0.1-1.6) 
for our sample galaxies. 
The left panel shows the SED fitting for one example of LSBGs 
(same as in Fig.~\ref{Fig2},  but the derived age is 2.10 Gyr), 
the middle panel shows the SED fitting for
one example of HSBGs (same as in Fig.~\ref{Fig2} but the derived age is 1.96 Gyr), and
the right panel shows the histogram
distributions of the derived ages of LSBGs (shade region) and HSBGs (dashed-line)
with bin of 0.2 Gyr.
}
   \label{Fig3}
\end{figure}

Figure~\ref{Fig3} shows the resulted SED fittings of two example galaxies
(the left panel for the same example of LSBGs, and the
middle panel for the same example of HSBGs as in Fig.~\ref{Fig2}) and 
the histogram distributions of the derived ages of the galaxies (the right panel,
where the shaded region is for LSBGs and the dashed-line is for HSBGs).
The right panel also shows that the ages of most of the LSBGs are 1-5 Gyr with median (mean)
value of 2.06 (2.18) Gyr, which means that the majority of their stars 
were formed quite recently. 
However, about 2\% of the LSBGs have larger ages,
5-8 Gyr, which mean that they could form their stars at earlier time.
Similarly, the ages of most of HSBGs are also 1-5 Gyr 
with the median value of 1.86 (mean as 1.92).
About 3\% of them have larger ages with 5-8 Gyr.
This similarity may mean that the LSBGs and HSBGs have 
no much different star formation history, but the HSBGs
are 0.2 Gyr younger than the LSBGs generally. 
All these results are consistent with those obtained with constant 
dust extinction,
although the varying $A_V$ results in a bit older age ($\sim$0.4 Gyr) 
than the constant one.
{The corresponding ages are also presented in Table~\ref{tab1}.}

\section{Incompleteness effects}

It is important to test the completeness of the sample,
and discuss its effect on the derived ages of the galaxies.

\subsection{The volume-limited sub-sample}

The incompleteness of the sample could affect the derived results of our
analysis, thus it is important to test. 
Zhong et al. (2008) has carefully discussed the incompleteness of the sample
of LSBGs.
In their Sect.~2, the good completeness of the SDSS MGS has been discussed.
The MGS is a spectroscopic sample selected from the SDSS
photometric data. The completeness for objects with spectroscopy
observations is high, exceeding 99\%, and the fraction of galaxies
eliminated by surface brightness cut is very small ($\sim0.1$\%).
Relative to all the SDSS targets, the SDSS spectroscopic survey is
90\% complete (Blanton et al. 2003; Hogg et al. 2004; Strauss et
al. 2002; McIntosh et al. 2006).
Blanton et al. (2005) had nicely discussed the incompleteness at low
surface brightness, and presented the contributions to completeness as a 
function of surface brightness, the $r$-band Petrosian half-light
surface brightness $\mu_{50,r}$.  It shows that, for
those brighter ones with $\mu_{50,r}<$23 mag arcsec$^{-2}$ 
(corresponding to $\mu_{0}$(B)=24.5 mag arcsec$^{-2}$), the
completeness of the spectroscopy is very close to 100\%.
But for
the photometric and tilling catalogs, they show obvious
incompleteness within $\mu_{50,r}$=22-23 mag arcsec$^{-2}$,
then the total completeness there could decrease to be about 70\%.

\begin{figure}
\centering 
\includegraphics [width=4.6cm, height=4.2cm]{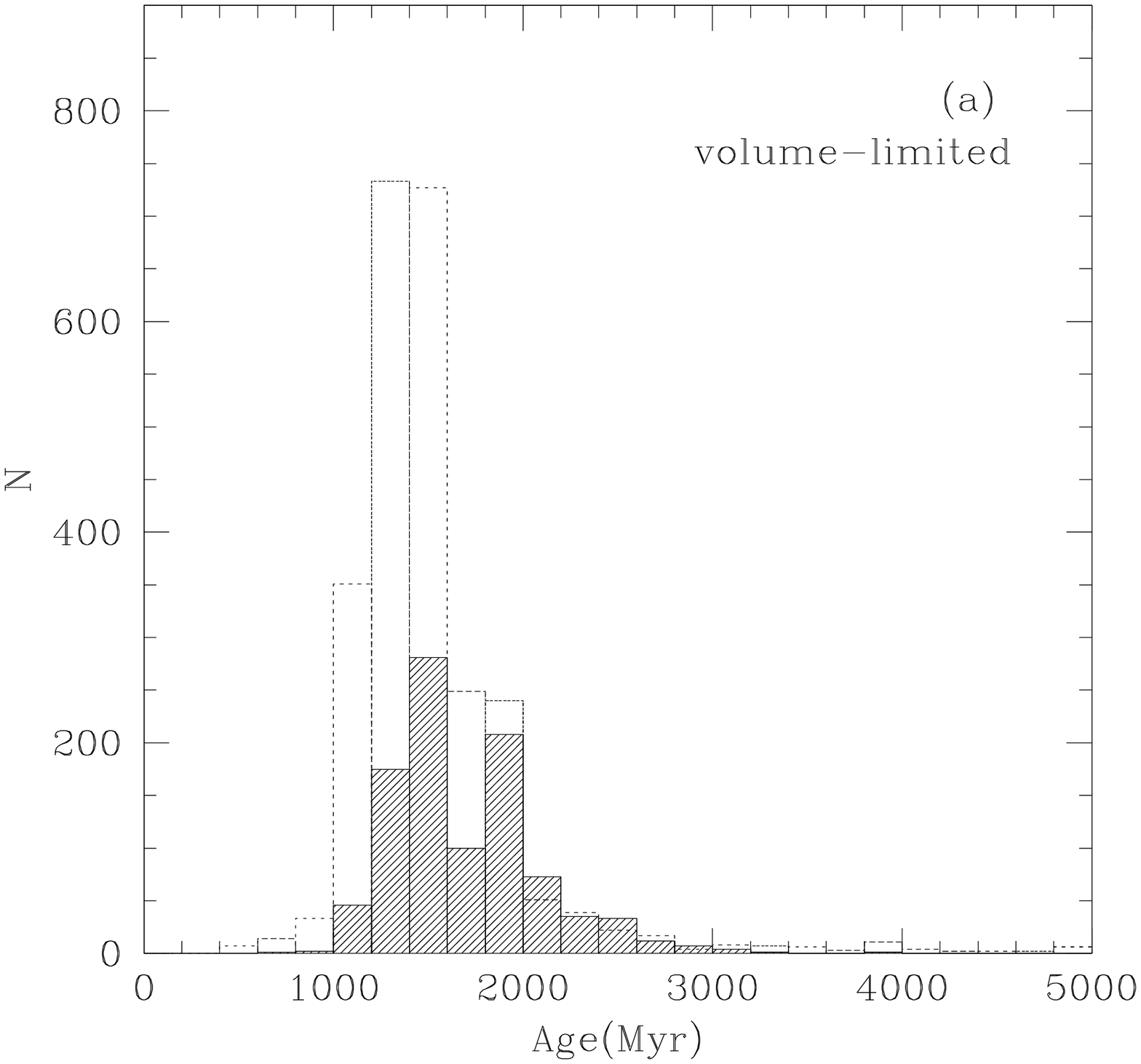}  \hspace{-0cm}
\includegraphics [width=4.6cm, height=4.2cm]{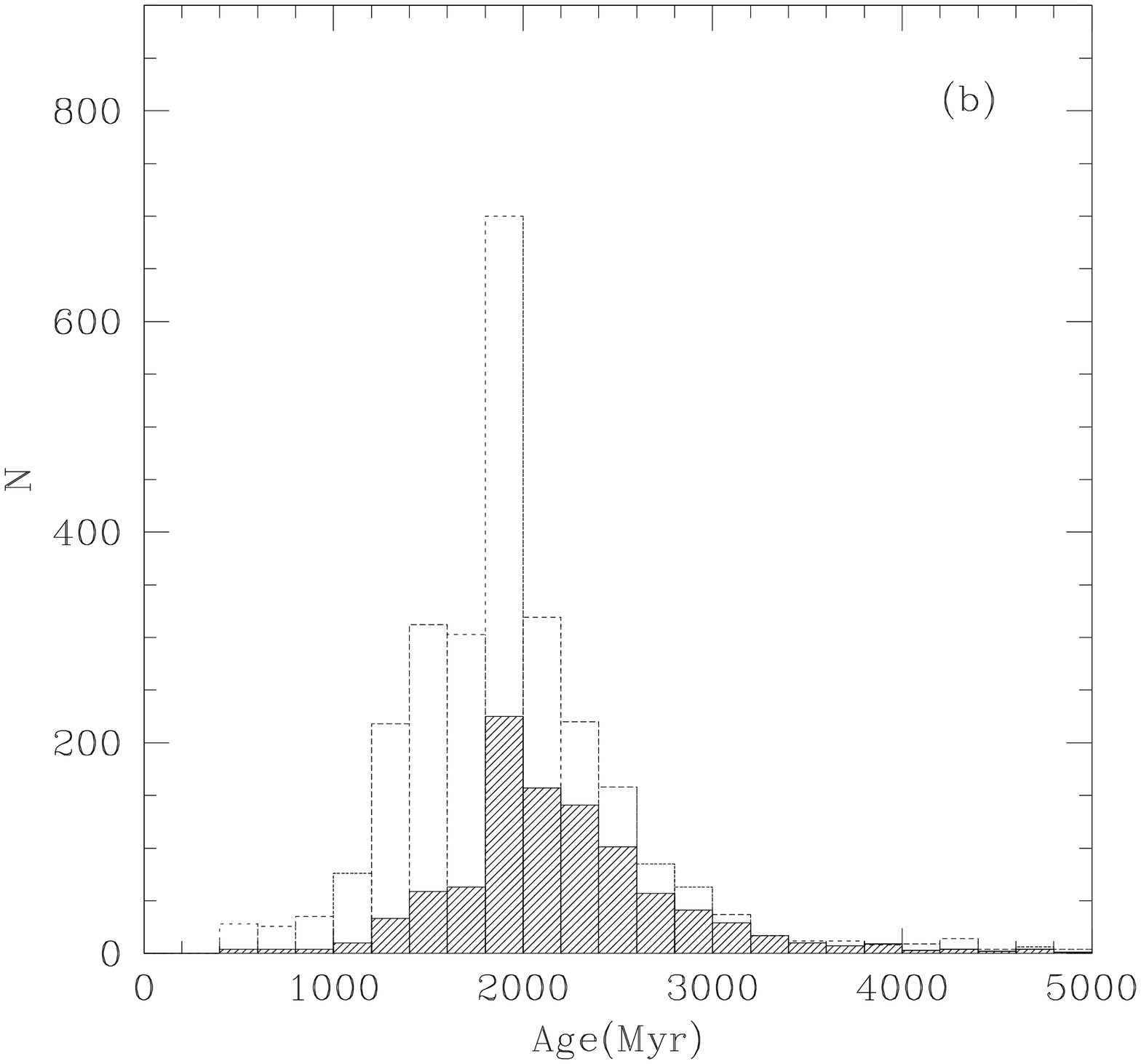} 
\caption{The histogram distributions of ages (bin as 0.2) of the volume-limited 
sub-samples of LSBGs (shade region) and HSBGs (dashed line).
The left panel is for the constant dust extinction case (see Sect.~4.2)
and the right panel is for the varying dust extinction case (see Sect.~4.3).
}
\label{fig.Mr.z}
\end{figure}

Similar to Zhong et al. (2008) in their Sect.5, here we also
extract a volume-limited sub-sample 
to minimize the effects of 
sample incompleteness on the derived ages of galaxies.
We firstly extract a volume-limited sub-sample from the $M_r-z$ plane 
 by considering $z<0.1$ and those brighter ones than the corresponding $M_r$
 from the SDSS-DR7 MGS as given in Sect.~\ref{sect:Obs}.1. 
 We obtain 5,608 LSBGs and 8,046 HSBGs.
Then, we match them with our sample galaxies 
having FUV to NIR SEDs (i.e., the 1,802 LSBGs and 5,886 HSBGs).
This allows us to obtain a sub-sample including 1,046 LSBGs and 2,853 HSBGs.
Their ages are given in Fig.~\ref{fig.Mr.z} as histogram distributions. 
The left panel is for the constant dust extinction, same as in Fig.~\ref{Fig2},
and the right panel is for the varying dust extinction, same as in Fig.~\ref{Fig3}.
The left one shows that
the median (mean) ages are 1.60 (1.69) Gyr for LSBGs and 1.44 (1.53) Gyr 
for HSBGs sub-samples.
The right panel shows that 
the corresponding median (mean) ages are 2.10 (2.20) Gyr for LSBGs 
and 1.90 (1.94) Gyr for HSBGs sub-samples.
These derived ages of the volume-limited sub-sample 
are very similar to those of the whole sample (Figs.~\ref{Fig2},\ref{Fig3}) 
with only 0.02-0.06 Gyr 
discrepancy in ages. This verifies the robustness of our results.
{The derived ages are also presented in Table~\ref{tab1}.}

\subsection{The pure disc sub-sample}
\label{sec.5.2}

\begin{figure}
\centering 
\includegraphics [width=4.6cm, height=4.2cm]{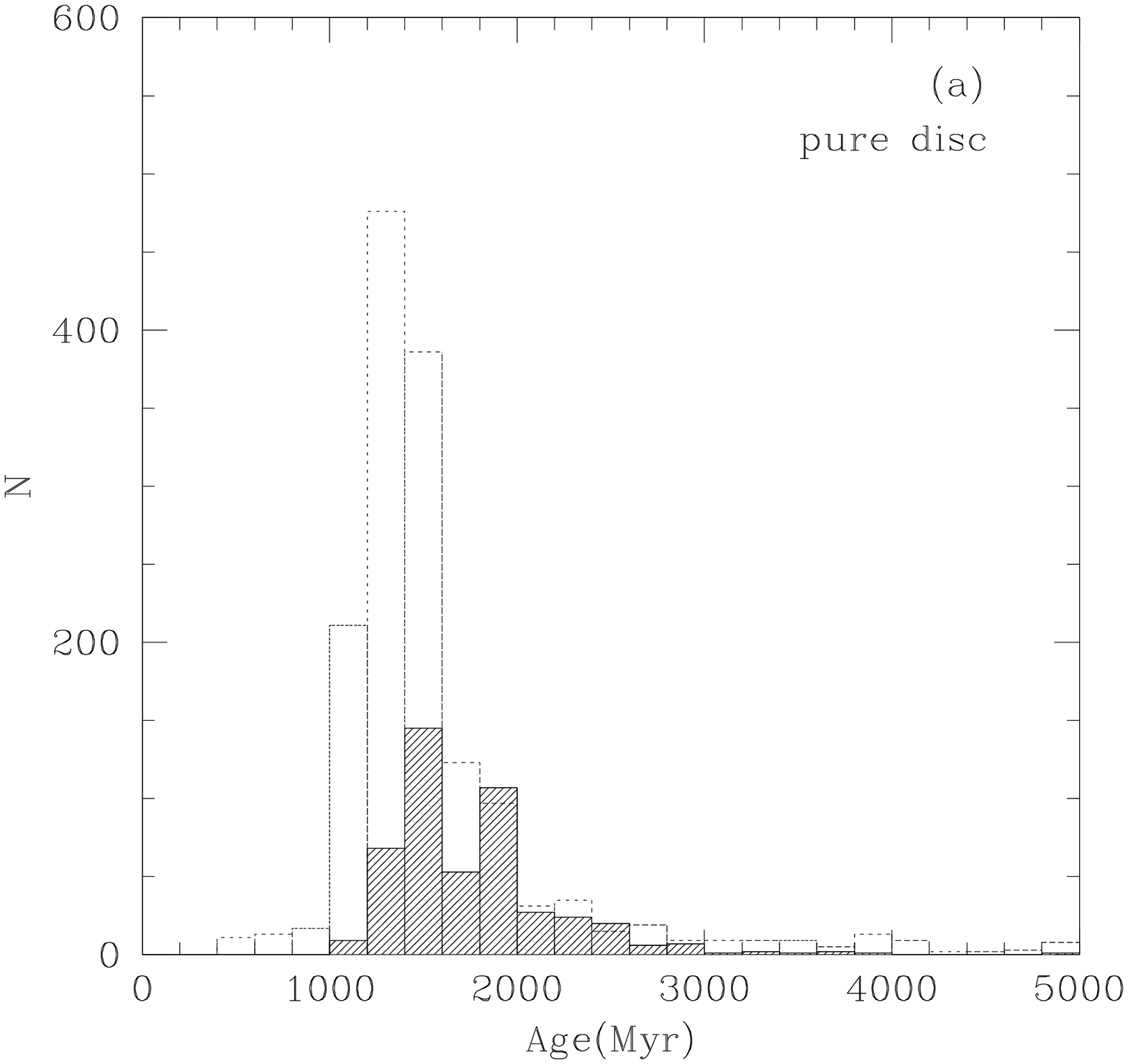}  \hspace{-0cm}
\includegraphics [width=4.6cm, height=4.2cm]{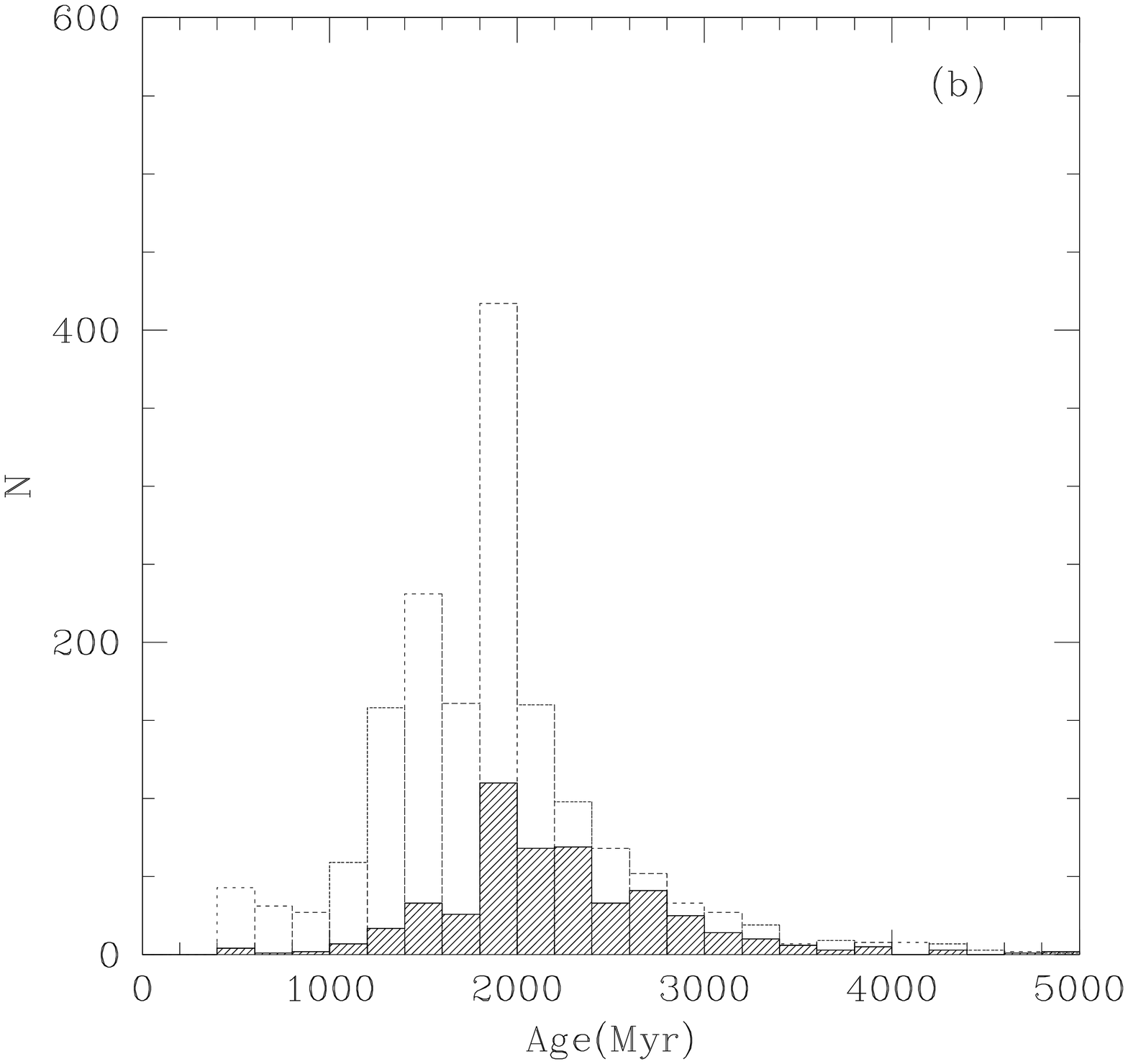} 
\caption{The histogram distributions of ages (bin as 0.2) of the 
pure disc sub-samples of LSBGs (shade region) and HSBGs (dashed line)
with $fracDev_r$ = 0.
The left panel is for the constant dust extinction case (see Sect.~4.2)
and the right panel is for the varying dust extinction case (see Sect.~4.3).
}
\label{fig.fraDev}
\end{figure}

 It is also necessary to check the effects of the calculated central surface brightness 
 on our results.  In the sample criteria (Sect.~2.1), we adopted $fracDev_r$ $<$ 0.25
 to guarantee the disk light (in r-band) to be well explained by exponential
profile. To entirely remove the effect of bulge light, we 
select a sub-sample as pure discs with $fracDev_r$ = 0.
This sub-sample includes 8,453 LSBGs and 12,096 HSBGs from the SDSS-DR7 MGS.
Then we match these sub-sample galaxies with our sample galaxies  
having FUV to NIR SEDs, then
we obtain 505 LSBGs and 1,818 HSBGs.  
The derived ages of these sub-sample galaxies are given in 
Fig.~\ref{fig.fraDev}.
The left panel shows the results with constant dust extinction,
and the median (mean) ages are 1.65 (1.77) Gyr for LSBGs and 
1.42 (1.58) Gyr for HSBGs sub-samples.
The right panel shows the results with varying dust extinction,
and the corresponding median (mean) ages are 2.10 (2.20) Gyr for LSBGs 
and 1.83 (1.87) Gyr for HSBGs sub-samples. 
These results are also very similar to the derived ages of the whole sample
of LSBGs and HSBGs (given in Fig.~\ref{Fig2}, Fig.~\ref{Fig3}), respectively,
as well similar to those of 
the volume-limited sub-sample (given in Fig.~\ref{fig.Mr.z}).
The discrepancy is 0.02-0.08 Gyr.
{These results are presented in Table~\ref{tab1} as well.}

We use criterion $fracDev_r<$0.25 to select the disc galaxies in our work
as shown in Sect.~2.1.
To carefully check whether the central surface brightness of galaxies depends on 
$fracDev_r$, we obtain the relations of $\mu_{0}$(B) vs. $fracDev_r$
for our LSBGs. It shows very small slope in their least-square fitting relation,
$-$0.08 (i.e., $\mu_{0}(B)=-0.08fracDev_r+22.51$). 
This negative value does show that, for the galaxies having
higher $fracDev_r$, their central surface brightness 
may be overestimated a bit by exponential disc profile, 
however, the discrepancy is quite small. We also check this 
relation for HSBGs, the corresponding slope is small as well, $-$0.30.
These verify that $fracDev_r<$0.25 could be a reasonable criterion 
to select disc galaxies, and won't affect much on our calculations on central 
surface brightnesses of galaxies. This criterion has already been more strict than
what used in literatures, for example, Chang et al. (2006) and Shao et al. (2007) 
used $fracDev_r<$0.5 to select their spiral galaxies.

\subsection{The $r \leq 16$ sub-sample}
\label{sec.5.3}

As we know 2MASS is shallow, thus
we may worry about the results obtained from 2MASS biased to 
the massive galaxies.
McIntosh et al. (2006) helpfully made cross-correlation on the well-defined and highly
complete spectroscopic selection of $r\leq 17.5$ mag galaxies in the SDSS MGS with
the 2MASS sources to explore the nature and completeness of the 2MASS (K-band) selection
of nearby galaxies.
They quantified the completeness of 2MASS galaxies in terms of optical properties from SDSS, and
found, for $r\leq 16$ mag, 94.5 per cent of the MGS is found in the 2MASS XSC.
An XSC completeness of 97.6 per cent is achievable at bright magnitudes, with blue
low-surface-brightness galaxies being the only major source of incompleteness. 
They 
concluded that the rapid drop in XSC completeness at $r > 16$ mag reflects the sharp
surface-brightness limit of the extended source detection algorithm in 2MASS.
A combined $K \leq 13.57$ and  $r\leq 16$ mag-limited selection provides the most 
representative inventory of galaxies in the local cosmos with NIR and optical measurements,
and 92.2 per cent completeness.

\begin{figure}
\centering 
\includegraphics [width=4.6cm, height=4.2cm]{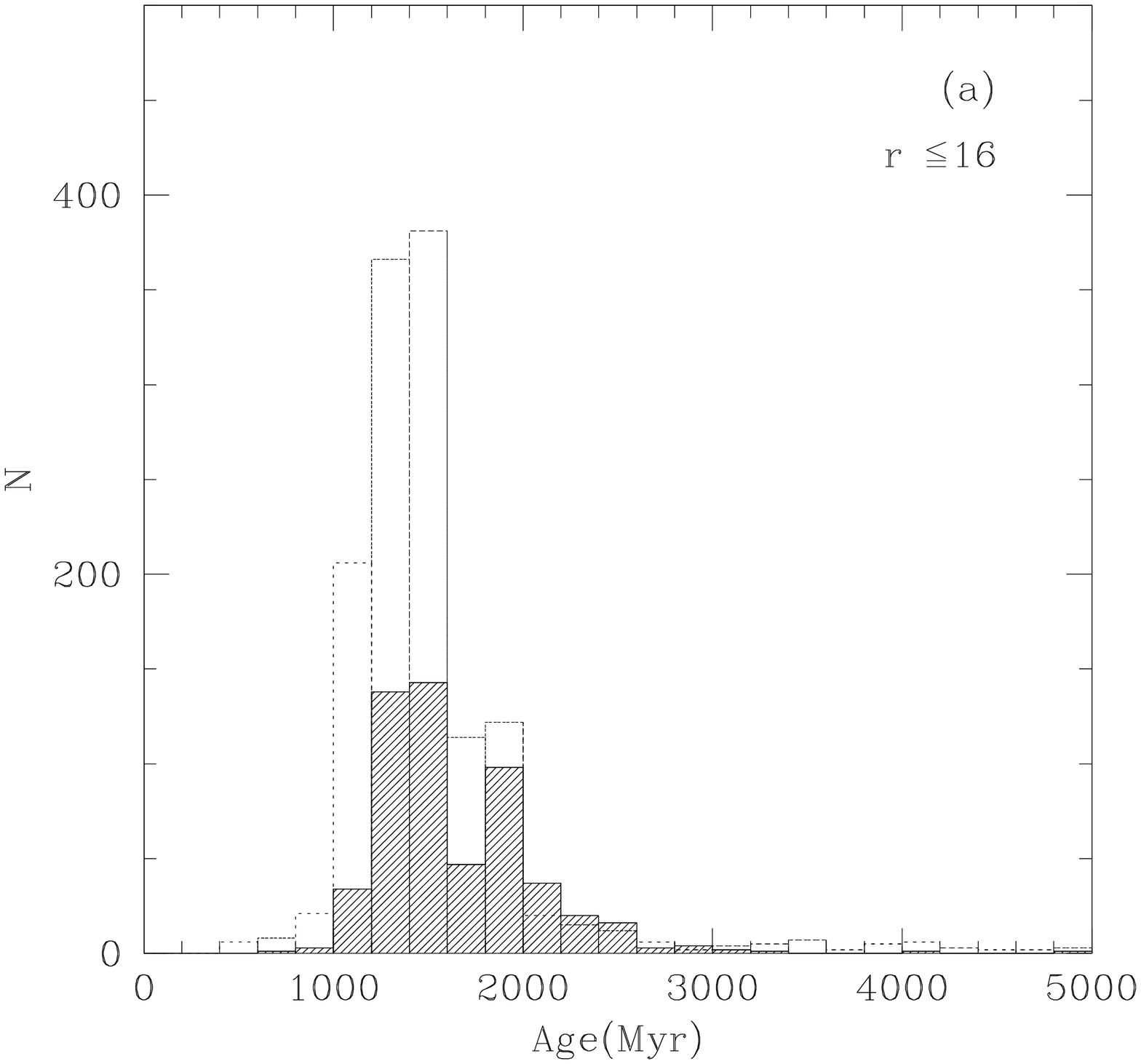}  \hspace{-0cm}
\includegraphics [width=4.6cm, height=4.2cm]{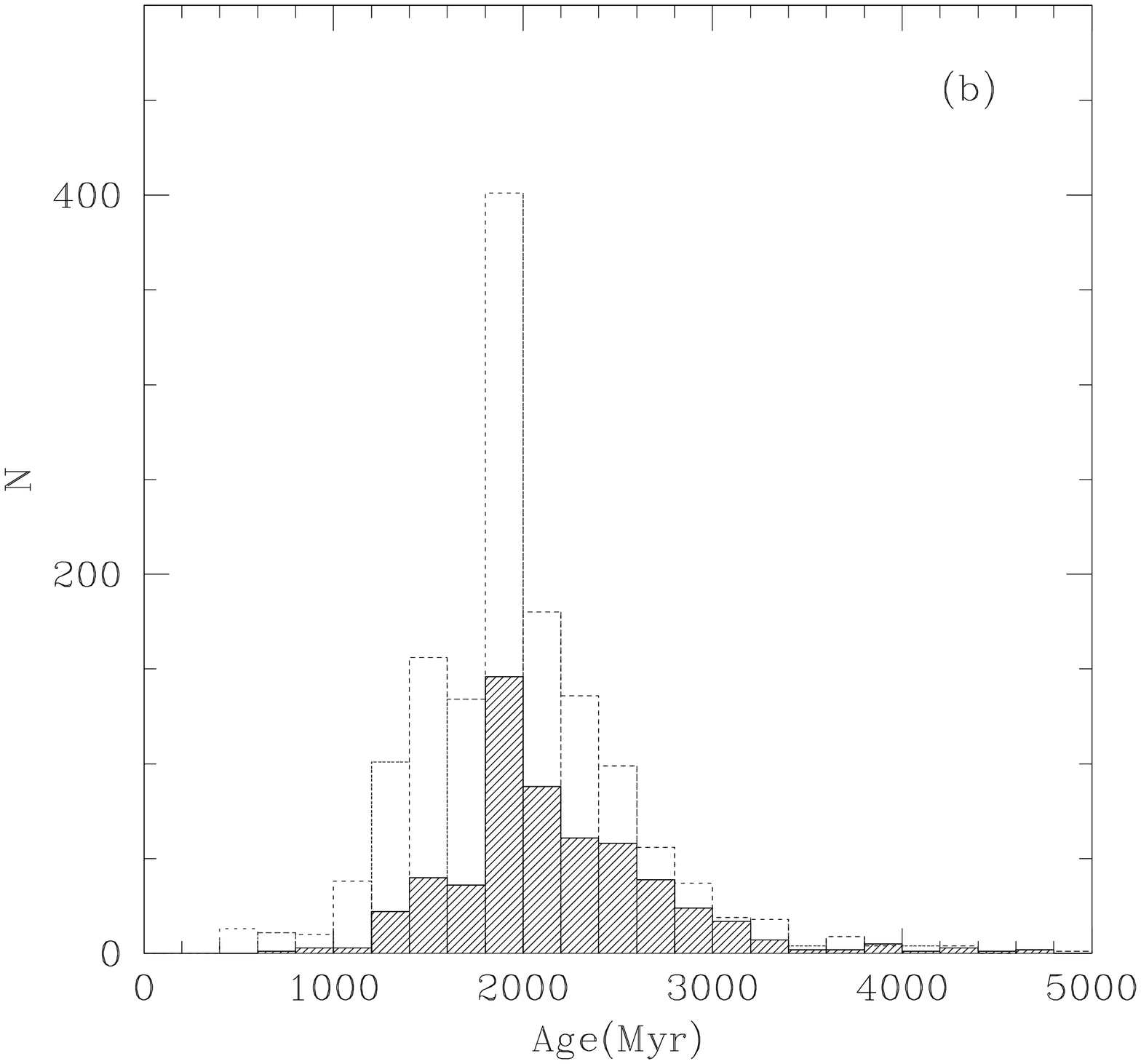} 
\caption{The histogram distributions of ages (bin as 0.2) of the 
$r \leq 16$ sub-samples of LSBGs (shade region) and HSBGs (dashed line).
The left panel is for the constant dust extinction case (see Sect.~4.2)
and the right panel is for the varying dust extinction case (see Sect.~4.3).}
\label{fig.r16}
\end{figure}

Therefore, we select a sub-sample with  $r\leq 16$ mag from
our LSBGs and HSBGs (they have $K \leq 13.57$), 
and then compare their derived ages with those of the whole sample.
Firstly, the galaxies with $r\leq 16$ are selected from the SDSS-DR7 MGS,
this results in 1,855 LSBGs and 2,485 HSBGs. These galaxies are further matched
with our sample galaxies having FUV to NIR SEDs, and then 
results in a sub-sample including 585 LSBGs and 1,553 HSBGs, respectively.
Their ages are given in Fig.~\ref{fig.r16} as histogram distributions. 
The left panel shows the results with constant dust extinction, and 
the median (mean) ages are 1.54 (1.65) Gyr for LSBGs and 1.42 (1.53) Gyr 
for HSBGs sub-samples. 
The right panel shows the results with varying dust extinction,
and the corresponding median (mean) ages are 2.07 (2.17) Gyr for LSBGs 
and 1.93 (1.98) Gyr for HSBGs
sub-samples.
These derived ages of the sub-sample are similar to those of
the whole sample as given in Fig.~\ref{Fig2} and Fig.~\ref{Fig3},
and to the other two sub-samples as given in Fig.~\ref{fig.Mr.z}
and Fig.~\ref{fig.fraDev}, respectively. The discrepancy is only up to 0.12 Gyr.
{These results are given in Table~\ref{tab1}.}

\subsection{The pure disc with $r \leq 16$ sub-sample}
\label{sec.5.4}

\begin{figure}
\centering 
\includegraphics [width=4.6cm, height=4.2cm]{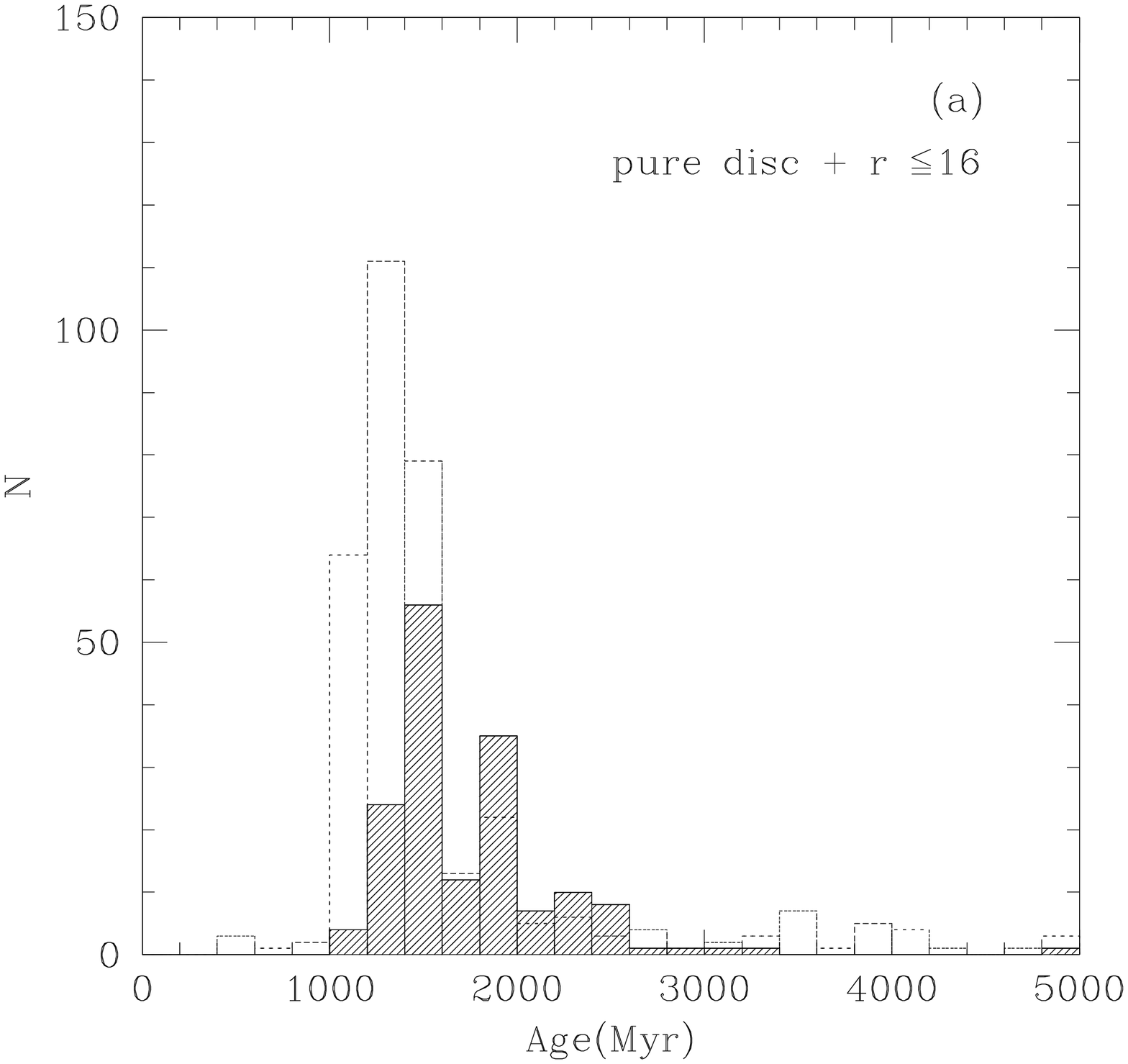}  \hspace{-0cm}
\includegraphics [width=4.6cm, height=4.2cm]{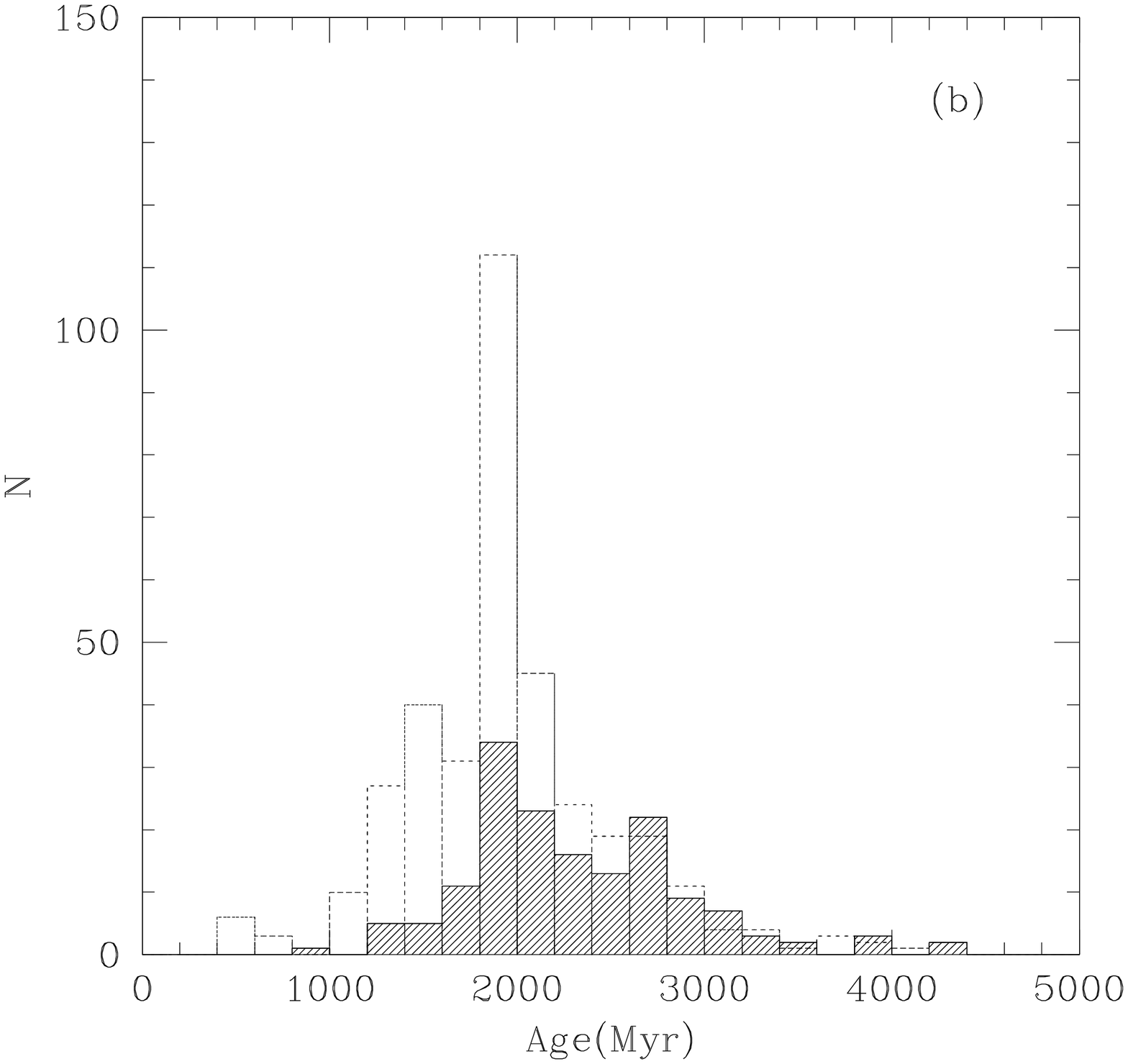} 
\caption{The histogram distributions of ages (bin as 0.2) of the 
pure disc with $r \leq 16$ sub-samples of LSBGs (shade region) and HSBGs (dashed line).
The left panel is for the constant dust extinction case
and the right panel is for the varying dust extinction case.}
\label{fig.frac0.r16}
\end{figure}

 In this subsection, we perform one more test to check the 
 robustness of our estimates by further extracting a subsample 
 with $fracDev_r$=0 and $r\leq 16$ mag, i.e. the pure discs with $r \leq 16$
 mag. That is, we select 
 the sub-sample by matching the galaxies in Sect.~\ref{sec.5.2}
 and Sect.~\ref{sec.5.3}. This results in 161 LSBGs and 341 HSBGs.
 The distributions of their derived ages are presented in Fig.~\ref{fig.frac0.r16}.
  
The left panel shows the results with constant dust extinction, and 
the median (mean) ages are 1.60 (1.74) Gyr for LSBGs and 1.39 (1.63) Gyr 
for HSBGs sub-samples. 
The right panel shows the results with varying dust extinction,
and the corresponding median (mean) ages are 2.17 (2.30) Gyr for LSBG
and 1.90 (1.97) Gyr for HSBG
sub-samples.
These derived ages of the sub-samples are similar to those of
the whole sample as given in Fig.~\ref{Fig2} and Fig.~\ref{Fig3},
and to the other three sub-samples as given in Figs.~\ref{fig.Mr.z},\ref{fig.fraDev},\ref{fig.r16}. 
The discrepancy is only up to 0.13 Gyr. 
The results are also given in Table~\ref{tab1}.

All the consistencies among the whole sample and the four sub-samples
about the derived ages of the sample galaxies confirm well
the robustness of our analyses. To be clear, all the results are summarized in Table~\ref{tab1}.

\begin{table} %[h!!!]
%\small
\centering
\bc
%\begin{minipage}[]{200mm}
\caption[]{\centering The derived ages (in unites of Gyr) of the sample and sub-sample galaxies. }
\label{tab1}
%\end{minipage}
\small
 \begin{tabular}{llllll}
  \hline\noalign{\smallskip}
              &  whole  & volume-limited    & pure disc &  $r \leq 16$  & pure disc + $r \leq 16$   \\
  \hline\noalign{\smallskip}
              &             &              &  LSBGs         &                &                 \\
Number        & 1,802       & 1,046        & 505            &  585           & 157 \\
Constant $A_V$ & 1.63 (1.75) & 1.60 (1.69) &  1.65 (1.77)    &   1.54 (1.65)  & 1.60 (1.74)             \\
Varying $A_V$  & 2.06 (2.18) & 2.10 (2.20) &  2.10 (2.20)    &   2.07 (2.17)  & 2.17 (2.30)      \\
  \noalign{\smallskip}\hline
              &             &              &  HSBGs         &                &                 \\
Number        & 5,886       & 2,853        & 1,818          &  1,553         & 392 \\
Constant $A_V$ & 1.47 (1.59) & 1.44 (1.53) &  1.42 (1.58)    &   1.42 (1.53)  & 1.39 (1.63)             \\
Varying $A_V$  & 1.86 (1.92) & 1.90 (1.94) &  1.83 (1.87)    &   1.93 (1.98)  & 1.90 (1.97)      \\
  \noalign{\smallskip}\hline  
\end{tabular}
\ec
\tablecomments{0.86\textwidth}{The first part is for the LSBGs
and the second part is for the HSBGs. ``Number" refers to the numbers of 
the galaxies in the sample and
sub-samples. The median (mean) ages (in unites of Gyr) of 
the galaxies are given for both of
constant and varying dust extinction cases.}
\end{table}

\section{Discussions}
\label{sect:discussion}

\subsection{Comparisons with previous studies}

 We study the ages of LSBGs using the EPS model PEGASE with exponential decreasing SFR 
 to fit their multiwavelength SEDs 
 from FUV to NIR. The HSBGs are also applied to similar studies as comparisons.
 Firstly, we find that the derived ages are
 1-5 Gyr for most of our large sample of 1,802 LSBGs, which is 
 consistent with most of the previous studies based on smaller sample. 
 
 Schombert et al. (2001) studied the V-I colors and relative
 H\,{\sc i} content of the most gas-rich LSB dwarf galaxies.
  They suggested that the low stellar densities of 
 their gas-rich LSB dwarfs are due to
 inefficient conversion of gas mass into stellar mass.
 The further comparison with star formation models (Boissier \& Prantzos 2000)
 indicated that the blue optical colors of LSB dwarfs  
 can only be explained by a 
 dominant stellar populations less than 5 Gyr in mean age.

  Zackrisson et al. (2005) used optical/NIR broadband photometry 
  together with H$\alpha$ emission line data to study the ages of
  a sample of 9 blue LSBGs. 
  They found that the current observations cannot rule out 
  the possibility that these blue LSBGs formed as recently as 1-2 Gyr ago. 
 Indeed, their Fig.3 shows that  the blue colors
of the group of blue LSBGs (B-V$\sim$
  0.4 mag, V-J$\sim$ 1.2 mag) can be well represented by an absolute age
  of 3.0 Gyr, an average age of 2.2 Gyr and a star formation history
  with $\tau$= 1.0 Gyr.  However, they also present possible much 
 older ages,
  such as 13 Gyr and 7.4 Gyr. Indeed, our results here from 
  more than thousands of LSBGs confirm their young age results. 

 In Vorobyov et al. (2009), 
 they used numerical hydrodynamic modelling and found 
  the existence of a minimum age for blue LSBGs:
  1.5-3.0 Gyr or 5-6 Gyr.  
  They complement hydrodynamic modelling with population synthesis modeling
  to produce the integrated B-V colors and H$\alpha$ equivalent widths.
  They adopted a sporadic model for star formation which yields no radial 
  abundance gradients in the model disk (as observed by
  de Blok \& van der Hulst 1998). 
  They also mentioned this sporadic star formation
  was agreement with the studies that the current star formation of LSBGs
  is localized to a handful of compact regions (Auld et al. 2006 from H$\alpha$
  imaging),
  and then there is little or no diffuse H$\alpha$ emission coming from the rest of
  the galactic disk. We may need more observations on LSBGs to further support this.   
  
 The most similar and comparable work with us is 
 Haberzettl, Bomans, Dettmar (2005), who studied the star formation 
 history of seven LSBGs in the HDF-S
 by comparing  the measured SEDs with the synthetic spectra extracted 
 from PEGASE. Comparing a library of SEDs to the measured spectra they were able to derive ages
between 2 to 5 Gyrs for the dominant stellar population. 
All these investigations above, as well as ours with a much larger sample,
favor a scenario that 
the major star formation events
of LSBG galaxies took place at much later stages (at z$\sim$0.2 to 0.4).

 However,
 Podoan et al. (1997) concluded that most of
 the LSBGs in their sample appeared to be older than about 7 Gyr. 
 In the work, Podoan et al. (1997) used a new IMF derived from numerical fluid dynamics
 simulations and a new synthetic stellar population code 
 obtained in their earlier works. Jimenez et al. (1998) further improved 
 their model and obtained similar results about ages of LSBGs (older than 7 Gyr).   
 In fact, a part of our LSBGs ($\sim$2-3\%) also have larger ages as 5-8 Gyr, 
 which could be consistent with Podoan et al. (1997) and Jimenez et al. (1998).
 These could mean that some of the LSBGs may do form their majority of 
 stars in earlier time.
 
 \subsection{Comparing the LSBGs and HSBGs}
 
 Another aspect of results in this work is that the HSBGs do not show 
 much different ages from the LSBGs. 
 The ages of most of the HSBGs are also 1-5 Gyr with part of them (3\%) have larger ages as
 5-8 Gyr.
 This is not unlikely and consistent with some other investigations.
Mattson, Caldwell, Bergvall (2007) study the chemical evolution of LSBGs. They conclude that 
LSBGs probably
have the same ages as their high surface brightness counterparts, although the global rate of star
formation must be considerably lower in these galaxies.
Boissier et al. (2003) compared the observed properties with the predictions of models of the chemical and
spectrophotometric evolution of LSBGs, the basic idea behind the models is that LSBGs are similar to
``classical" HSB spirals except for a larger angular momentum.

As van den Hoek et al. (2000) discussed, the presence of an old stellar 
population in many late-type LSBGs as indicated by the optical colors
and confirmed by their galactic chemical and photometric 
evolution model suggested that LSBGs roughly follow
the same evolutionary history as HSBGs, but at a much lower rate. 
The mean age of the stellar population in most LSBGs and HSBGs was similar 
even though the disks of LSBGs were in a relatively early evolutionary stage.

Although Haberzettl, Bomans, Dettmar (2005) found the HSBGs from Kennicutt
(1992) have much larger ages than their LSBGs, in fact, their Fig.2 shows that
some of the objects of Kennicutt (1992) do have quite young ages $<$4 Gyr, and
many of the objects of Kennicutt (1992) have SEDs as early-type galaxies (E or
S0), which should not be very similar objects to our galaxies here as face-on disks. 
Similarly, the relatively high ages of the HSBGs 
in Terlevich \& Forbes (2002) and Caldwell et al. (2003)
could also be due to most of their samples as early-type galaxies.

 The similar ages between LSBGs and HSBGs obtained in this work
could be consistent with 
  Liang et al. (2010) about the metallicity analysis of the LSBGs
  and HSBGs, which also come from the parent sample of Zhong et al. (2008). 
  They found that the LSBGs and HSBGs 
  located closely on the stellar mass vs. metallicity and N/O vs. O/H relations of the
normal galaxies, but the LSBGs have slightly higher N/O than the HSBGs 
at give O/H in low metallicity region. This 
 may mean that the LSBGs could have relatively lower SFR than the HSBGs 
and then show dominant primary nitrogen component there as Molla et al. (2006, their Fig.5) suggested. 
However, we should notice that the derived ages of HSBGs in this work are about 0.2 Gyr younger than those of the LSBGs
generally (see Table~\ref{tab1}). 
This may mean that HSBGs may have occurred star formation process
more recently than the LSBGs, 
or the recent star formation process is stronger in HSBGs than in LSBGs. 
This is also consistent with the results of Chen et al. (2010, in preparation)
for the stellar population analysis on spectral absorption lines
and continua of Liang et al (2010)'s sample, in which they found that the HSBGs have slightly larger fraction (5\%) 
of young population than the LSBGs. 
In a word, as Liang et al. (2010) concluded,   
the large sample shows that LSBGs span a wide range in metallicity
and stellar mass, and they 
lie nearly on the stellar mass vs. metallicity and N/O vs. O/H relations of
normal galaxies. The HSBGs show similar trends. These suggest that LSBGs and HSBGs 
 have not had dramatically different star formation and chemical 
 enrichment histories.

\subsection{Effects of other parameters}

We adopt the exponential decreasing SFR in PEGASE
for fitting the SEDs of the sample galaxies.
The derived star formation rate decay time $\tau$ of LSBGs spread in a wide range from 0.1 to 15 Gyr
in both cases of constant (Sect.\,4.1) and varying (Sect.\,4.2) dust extinction.
In the constant $A_V$ case the median of $\tau$ is about 0.5 Gyr 
and mean as 0.57 Gyr. 
In the varying $A_V$ case, the median of $\tau$ is about 0.5 Gyr 
and mean as 0.62 Gyr. 
For the HSBGs, the derived $\tau$ values 
are also within 0.1-15 Gyr with median (mean) of 0.5 Gyr (0.78 Gyr) 
in constant $A_V$ case, and
median (mean) of 0.6 Gyr (0.73 Gyr) in varying $A_V$ case, respectively.
It is similar between LSBGs and HSBGs. 
These values are not unacceptable.
Haberzettl, Bomans, Dettmar (2005) estimated $\tau$
to be 500 Myr (2 cases), 1400 Myr (3 cases), 5000 Myr (2 cases) for their 7 LSBGs.
Li et al. (2004a) adopted $\tau$ = 3 Gyr for M81,
and Li et al. (2004b) adopted $\tau$ = 12 Gyr for M33 in their 
SED fittings by using PEGASE. Zackrisson et al. (2005) also
adopted wide range of $\tau$ in their model (0.5-15 Gyr generally, their Fig.3).

The colors and H$\alpha$ emission properties of disk and irregular galaxies
have shown the general picture of their star formation history. 
As Kennicutt (1983) and Kennicutt et al. (1994) commented, 
early-type galaxies (types S0-Sb) represent systems which formed most of their
gas into stars on timescales much less then the Hubble time, while the disks of
late-type systems (Sc-Im) have formed stars at roughly a constant rate 
since they formed. They parameterized the star formation history as 
an exponentially declining star formation rate:
{\it SFR(t)} = $R_0$ {\it e}$^{-t/\tau}$, where $\tau$ was adopted as 0-15 Gyr,
and the 1/$\tau$=0 case is corresponding to the constant star formation case (Kennicutt 1983). 
As Zackrisson et al. (2005) analyzed, the short burst scenario (with star formation
ending very abruptly) can be ruled out for LSBGs since 
it will predict too high EW(H$\alpha$). And in the scenarios including constant or increasing
star formation rates over cosmological time scales, it still predict
too high EW(H$\alpha$) to reach the observations. This may nonetheless be remedied if the 
slope of the IMF is significantly more bottom-heavy or the upper mass limit 
substantially lower than typically assumed.

For LSBGs, Vorobyov et al. (2009) used a simple 
model of sporadic star formation in which the individual star formation sites (SFSs) are
distributed randomly throughout the galactic disk, which is reasonable for LSBGs.
They further point out, there is evidence that star formation in blue LSBGs does
not proceed at a near-constant rate. Their own numerical simulations and modelling by 
Zackrisson et al. (2005)
indicated that the SFR should be declining with time to reproduce the observed EW(H$\alpha$)
in LSBGs. Therefore, they assumed an exponentially decreasing SFR in their modelling.
Moreover, van den Hoek et al. (2000) found that most of the LSBGs in their sample
belonged to the group of late-type galaxies for which exponentially decreasing SFR models were in 
good agreement with the observations. 
Haberzettl, Bomans, Dettmar (2005) also pointed out that 
the spectra of their sample galaxies were all represented by an exponential
decreasing star formation rate.
Therefore, it is reasonable for us to adopt the exponential
decreasing SFR for our LSB sample galaxies by following the discussions above.
We also adopt the same form of SFR for the HSBGs to be consistent. 
Even though, 
we have tried to use the constant SFR instead in PEGASE to fit the SEDs of all our sample
galaxies. The resulted ages of galaxies become a bit larger than those with exponential decreasing SFR.
It is $\sim$5-6 Gyr generally. This could be understood from the model prediction,
i.e., for the constant SFR case, it may need longer time to assemble the equivalent amount 
of stellar populations.

The degeneracy between age and metallicity is often a problem in 
stellar population analysis of galaxies. It is not easy to degenerate them.
In our fittings, we obtain the  metallicity Z of galaxies from  0.0001 to 0.04 in a wide range
for most of the sample galaxies.
As we know, the metallicity presented by PEGASE is the value averaged on the Simple Stellar
Populations (SSPs) with various
metallicities, and the average is sensitive to the dominant populations among all the SSPs.
Moreover, by adding UV and NIR photometric data to the optical data,
the degeneracy between age and metallicity could be efficiently broken
as discussed in Sect.~4.1.
Thus the dominate stellar population ages derived from 
PEGASE on SED fittings here should be robust.

The derived ages of galaxies discussed here represent 
the time-scale since their dominant stellar populations formed. 
This could be different from the stellar mean ages of the galaxies. 
PEGASE model also provides the mean ages of the stars 
averaged on the bolometric luminosity. 
We check their values by following the items in Table\,1, 
and found that these mean ages of stars are about 0.9 Gyr younger generally.
This could be understood from the model prediction since the stars are assumed to form later than that time
corresponding to the ages of galaxies.

\section{Conclusions}
\label{sect:conclusion}

  We summarize and conclude our work in four items.
  
  1) A much large sample (1,802) of nearly face-on disk LSBGs are studied their ages by fitting their multiwavelength
  SEDs from FUV to NIR using the EPS model PEGASE. The exponential decreasing SFR is adopted. 
  The ages of LSBGs spread in a wide range.
  Most of them have ages of 1-5 Gyr generally no matter the constant
  or varying dust extinction $A_V$ are considered 
  (the varying dust extinction results in $\sim$0.4 Gyr older ages than the
  constant one). 
  This age range is consistent with most of previous studies on smaller sample of LSBGs.
  In addition, a part of the LSBGs ($\sim$2-3\%) have larger ages as 5-8 Gyrs.
 These derived ages are also consistent with some earlier works.  
  We should notice that if a constant SFR is adopted instead, the derived ages
 of sample galaxies will become a bit larger, $\sim$5-6 Gyr generally, which is for
 a further check although it has been commented that the exponential decreasing SFR
 is favored for LSBGs.
  
  2) 
  A large sample (5,886) of nearly face-on disk HSBGs are also selected and 
  studied using same procedure for comparisons.  
  The results show that most of these HSBGs also have ages 
  1-5 Gyr, but with $\sim$3\% having larger ages as 5-8 Gyrs, 
  which are not much different from the LSBGs. However, the HSBGs are about 0.2 Gyr younger
  than the LSBGs, which may 
 indicate that the HSBGs have more recent star forming activities than the LSBGs. 
 
 3)
  Four sub-samples are further selected for checking the incompleteness effects:
  the volume-limited one selected from the $M_r-z$ plane ($z<0.1$ and 
  those brighter ones than the corresponding $M_r$),
  the pure disc one with $fracDev_r$=0, the  $r \leq 16$ one for testing 
  2MASS completeness, and further the pure disc with $r \leq 16$ one. 
  All the four sub-samples show quite similar 
  results to the whole sample, i.e., the derived ages of the sample galaxies
  are quite similar to each other, and only have small discrepancy, 0.02-0.13 Gyr.

 4)
  The similar ages between LSBGs and HSBGs
 could be consistent with the metallicity and stellar population analysis of them
 (Liang et al. 2010; Chen et al. 2010, in preparation).
 These suggest that LSBGs and HSBGs 
 have not had dramatically different star formation and chemical 
 enrichment histories.

\normalem
\begin{acknowledgements}
We thank our referee for the very valuable comments and suggestions, 
and very efficient reviewing, 
which help in improving well our work. 
We also thank our Editor, Prof. Changbom Park, for his very efficient 
management on our paper.
 We thank Philippe Prugniel, Zhengyi Shao, Ruixiang Chang, Hector Flores, Myriam Rodrigues, Mathieu Puech, Chantal
 Balkowski, Rodney Delgado, Sylvain Fouquet, Jianling Wang
 and Xuhui Han for helpful discussions.
 This work was supported by the Natural Science Foundation of China
 (NSFC) Foundation under Nos.10933001, 10973006,
10973015, 10673002; the
 National Basic Research Program of China (973 Program) Nos.2007CB815404,06;
 and the Young Researcher Grant of National Astronomical
Observatories, Chinese Academy of Sciences.
 We thank the wonderful SDSS, 2MASS and GALEX database,
 and the wonderful NYU-VAGC, CASJobs and MPA/JHU/SDSS.
 
\end{acknowledgements}

\label{lastpage}

\end{document}